\documentclass{article}
\usepackage{multirow}
\usepackage[left=20mm,right=20mm,top=20mm,bottom=20mm]{geometry}

\usepackage{natbib}
\usepackage{amssymb,amsmath}
\usepackage{color}
\usepackage{eurosym}
\usepackage{graphicx}
\usepackage{longtable}
\usepackage{multirow}
\usepackage{array,color,colortbl}
\usepackage{booktabs}
\usepackage[dvipsnames]{xcolor}
\usepackage{hyperref}
\usepackage[width=\textwidth]{caption}
\usepackage{textcomp}
\usepackage{pdflscape}
\usepackage{colortbl}

\title{Interior Dynamics and Thermal Evolution of Mars -- a Geodynamic Perspective}

\author{\normalfont Ana-Catalina Plesa$^1$, Mark Wieczorek$^2$, Martin Knapmeyer$^{1}$, \\ Attilio Rivoldini$^3$, Michaela Walterova$^1$, Doris Breuer$^1$}

\date{}

\begin{document}

\maketitle
\begin{center}
$^1$German Aerospace Center (DLR), Berlin, Germany.\\
$^2$Universit\'{e} C\^{o}te d'Azur, Observatoire de la C\^{o}te d'Azur, CNRS, Laboratoire Lagrange, Nice, France.\\
$^3$Royal Observatory of Belgium, Brussels, Belgium.
\end{center}

\begin{abstract}
Over the past decades, global geodynamic models have been used to investigate
the thermal evolution of terrestrial planets. With the increase of computational
power and improvement of numerical techniques, these models have become
more complex, and simulations are now able to use a high resolution 3D
spherical shell geometry and to account for strongly varying viscosity,
as appropriate for mantle materials. In this study we review global 3D
geodynamic models that have been used to study the thermal evolution and
interior dynamics of Mars. We discuss how these models can be combined
with local and global observations to constrain the planet's thermal history.
In particular, we use the recent InSight estimates of the crustal thickness,
upper mantle structure, and core size to show how these constraints can
be combined with 3D geodynamic models to improve our understanding of the
interior dynamics, present-day thermal state and temperature variations
in the interior of Mars. Our results show that the crustal thickness variations
control the surface heat flow and the elastic thickness pattern, as well
as the location of melting zones in the present-day martian mantle. The
lithospheric temperature and the seismic velocities pattern in the shallow
mantle reflect the crustal thickness pattern. The large size of the martian
core leads to a smaller scale convection pattern in the mantle than previously
suggested. Strong mantle plumes that produce melt up to recent times become
focused in Tharsis and Elysium, while weaker plumes are distributed throughout
the mantle. The thickness of the seismogenic layer, where seismic events
can occur, can be used to discriminate between geodynamic models, if the
source depth and location of seismic events is known. Furthermore model
predictions of present-day martian seismicity can be compared to the values
measured by InSight. Future models need to consider recent estimates from
the present-day elastic lithosphere thickness at the north pole of Mars,
the effects of lateral variations of seismic velocities on waves propagation
through the mantle and lithosphere, and to test the spatial distribution
of seismicity by comparing model predictions to observations.

\end{abstract}


\section{Introduction}
\label{sec:Intro}

A variety of numerical modeling studies have been used to investigate the
thermal evolution and interior dynamics of Mars
\citep[see][for a review]{breuer15,smrekar19}. These studies use either
fully dynamical 2D/3D simulations that self-consistently model the evolution
of mantle flow and through their nature can address local output quantities
in regions of interest or 1D parametrized models that track the evolution
of global quantities, such as the global mantle temperature or the average
surface heat flow. Fully dynamical 2D and 3D models have been used to predict
the surface heat flow and seismicity distribution on present-day Mars
\citep{plesa16,plesa18}, the effects of impacts during the early thermochemical
history \citep{ruedas17,roberts17}, partial melting and crust-mantle differentiation
\citep{ruedas13,plesa14b}, and the formation of the martian crustal thickness
dichotomy \citep{keller09,golabek11}. On the other hand, 1D parametrized
thermal evolution models have been applied to investigate the crustal formation
and crust-mantle differentiation
\citep{hauck02,breuer06,morschhauser11}, as well as mantle degassing
\citep{fraeman10,grott11}, the magnetic field evolution
\citep{breuer03,williams04}, the thermal state of the lithosphere
\citep{grott08,thiriet18}, and the coupled thermal-orbital evolution of
Mars and its moon Phobos \citep{samuel19}.

With the arrival of the InSight mission in 2018 \citep{banerdt20}, a new
chapter of planetary geophysics has begun. Equipped with a seismometer
(SEIS, \citealp{lognonne20}), a heat flow probe (HP$^{3}$,
\citealp{spohn18}) and X-band telecommunication capabilities that allow to
precisely measure the rotation of Mars (RISE, \citealp{folkner18}), InSight
investigates the interior of Mars from its core to the surface.

InSight's measurements represent the most direct set of constraints for
the interior of Mars. The crustal thickness
\citep{knapmeyer-endrun21}, the thickness and thermal state of the lithosphere
\citep{khan21}, and the size of the core \citep{staehler21} provide important
information for modeling of thermal evolution and investigating physical
processes active in the interior of Mars. In turn, numerical models of
the thermal evolution combined with InSight measurements can be used to
constrain poorly known parameters, such as the rheology of the mantle,
distribution of heat producing elements (HPEs) between the mantle and crust,
and the evolution of the surface and core-mantle boundary (CMB) heat flows.
Previous 1D models that used inversion techniques for the interior structure,
composition, and thermal state of the mantle or made assumptions of the
present-day thermal structure have made predictions about the seismic structure
of the martian mantle
\citep[e.g.,][]{khan17,zheng15,gudkova04,sohl97,mocquet96}. More recent
thermal evolution models have directly addressed the seismic observations
of InSight. These models investigated the effects of crustal thickness
and its enrichment in HPEs on the thermal evolution and present-day partial
melt distribution in the interior of Mars \citep{knapmeyer-endrun21}, studied
the consequences of a molten layer at the base of the mantle on the thermal
history and core size measurements \citep{samuel21}, and estimated seismic
velocities variations due to the interior temperature distribution
\citep{plesa21}.

In this study we focus on 3D thermal evolution models and discuss results
presented in previous studies and new simulations in the framework of the
recent InSight data about the martian crust, mantle, and core. In
Section~\ref{sec:Geodyn} we review the mathematical equations used by
geodynamic models to calculate the thermal evolution of Mars.
Section~\ref{sec:Lith} presents the effects of the crustal thickness
variations on model predictions for heat flow and elastic lithospheric
thickness variations, as well as on the distribution of partial melt zones in
the mantle. The dynamics inside the martian mantle, seismic velocities
variations due to mantle thermal anomalies, and constraints from tidal
deformation are discussed in Section~\ref{sec:MantleDyn}. In
Section~\ref{sec:Core} we list the core size estimates from geodetic data
and seismic observations and discuss their consequence on the mantle dynamics
and convection pattern. The seismogenic layer thickness and the present-day
seismicity are discussed in Section~\ref{sec:Seismicity}. In the last
section (Section~\ref{sec:Conclusions}) we present a summary of
geodynamic models and their findings and suggest future investigations.

We discuss in detail selected models that are compatible with the crustal
thickness and core radius estimates that were derived from InSight's seismic
measurements. Throughout this study we illustrate how model results can
be related to available geological and geophysical observations. For our
models we use both global constraints such as the thermal lithosphere thickness
\citep{khan21} and estimates of the tidal deformation of Mars
\citep{genova16,konopliv16,konopliv20}, as well as local data sets, such
as the present-day elastic lithosphere thicknesses at the north and south
poles \citep{phillips08,wieczorek08,broquet20,broquet21} and the location
of partial melt that could explain recent volcanic activity in Tharsis
and Elysium volcanic centers. We highlight how our knowledge of the thermal
history and present-day state of the interior has improved with the recent
InSight results and discuss what future models need to address in order
to further constrain the thermal evolution of Mars. Current geodynamic
modeling will also help to identify scientifically interesting landing
sites for a future seismic network that may become available after 2030
\citep{staehler22}.

\section{Geodynamic modeling and thermal evolution of Mars}
\label{sec:Geodyn}

Geodynamic models that are used to investigate the interior dynamics of
rocky planets in general, and of Mars in particular, numerically solve
for a set of conservation equations. While a 3D spherical geometry is the
choice for investigating the convection pattern in the interior and relating
the results to local observations, 2D models have the advantage of being
computationally faster making it possible to test a larger parameter space
with a higher spatial resolution compared to their 3D counterparts. The
equations and parameters used in these models are typically nondimensional.
They are scaled with the mantle thickness $D$ as length scale, a reference
thermal diffusivity $\kappa $ as time scale, and the initial temperature
difference $\Delta T$ across the mantle as temperature scale, although
some codes use dimensional quantities \citep{kronbichler12,heister17}. The advantage of the nondimensionalization
is that characteristic dimensionless numbers such as the Rayleigh number,
which is defined as the ratio between parameters driving convection and
those opposing it, can be used to describe the convection system without
needing to know the exact detailed parameters of a simulation.

The system of equations that is solved includes the conservation of mass,
linear momentum and thermal energy \citep[e.g.,][]{schubert01}. Their nondimensional
formulation for a system assuming a Newtonian rheology, an infinite Prandtl
number as appropriate for high viscosity media with negligible inertia,
a variable thermal expansivity and conductivity, using the Extended Boussinesq
Approximation (EBA), and including solid-solid phase transitions read
\citep[e.g.,][]{christensen85}:
%
\begin{align}\label{eqns}
\textstyle \nabla \cdot \vec{u} & = & 0, \\ 
\textstyle \nabla \cdot \left[\eta(\nabla \vec{u}  + (\nabla \vec{u})^T)\right] - \nabla p  + (Ra \alpha T - \sum_{l=1}^3Rb_l\Gamma_l)\vec{e}_r & = & 0, \\ 
\textstyle \frac{D T}{D t} - \nabla \cdot (k \nabla T) - Di \alpha (T+T_0)u_r - \frac{Di}{Ra}\Phi &  & \nonumber \\ 
- \textstyle \sum_{l=1}^3Di\frac{Rb_l}{Ra}\frac{D\Gamma_l}{Dt}\gamma_l(T+T_0) - H & = & 0, 
\end{align}
where $\vec{u}$ is the velocity vector, $u_{r}$ is its radial component,
$\eta $ is the viscosity, $p$ is the dynamic pressure, $\alpha $ is the
thermal expansivity, $T$ is the temperature, $\vec{e}_{r}$ is the unit
vector in radial direction, $t$ is the time, $k$ is the thermal conductivity,
$Di$ is the dissipation number, and
$\Phi \equiv \underline{\tau} : \underline{\dot{\varepsilon}} / 2$ is the
viscous dissipation, where $\underline{\tau}$ and
$\underline{\dot{\varepsilon}}$ are the deviatoric stress and strain-rate
tensors, respectively.

The Rayleigh number that describes the vigor of convection, the internal
heating rate $H$ that controls the mantle heating due to radioactive elements
(HPEs), and the dissipation number $Di$ that accounts for the increase
of temperature due to adiabatic compression effects are defined as follows:
%
\begin{equation}\label{eq_Rayleigh}
Ra = \frac{\rho g \alpha \Delta T D^3}{\eta \kappa}, \hspace{5mm} 
H = \frac{\rho Q_{HPE} D^2}{k \Delta T}, \hspace{5mm} 
Di=\frac{\alpha g D}{c_p},
\end{equation}
where $g$ is the gravitational acceleration, $c_{p}$ is the mantle heat
capacity, and $Q_{HPE}$ is the heat production rate in W\,kg$^{-1}$.

Previous geodynamic models accounted for two exothermic phase transitions,
and, in case of a small core radius, for an additional endothermic phase
change. While exothermic phase transitions tend to accelerate mantle flow,
the endothermic one slows it down and promotes layered convection. In particular,
the endothermic phase transition from wadsleyite/ringwoodite to bridgmanite
has been suggested to affect the style of convection. This phase transition
significantly changes the CMB heat flow and leads to a low degree convection 
pattern that was proposed by previous studies to explain the formation
of the Tharsis volcanic province
\citep{harder96,breuer98,spohn98,vanThienen06}. However, this phase
transition is not relevant for models with a core radius of 1700 km or
larger, and in these scenarios the convection in the mantle is characterized
by more than one plume \citep{spohn98,michel11}.

One of the most important parameters in geodynamic models is the mantle
viscosity, which is temperature and pressure dependent and follows an Arrhenius
law. Its nondimensional formulation \citep[e.g.,][]{roberts06} reads:
%
\begin{equation}\label{eq_viscosity}
\eta(T,z) = \exp\left(\frac{E+zV}{T+T_0} - \frac{E+z_{ref}V}{T_{ref}+T_0}\right),
\end{equation}
where $z$ is the depth, $E$ and $V$ are the activation energy and activation
volume, respectively, $T_{0}$ is the nondimensional surface temperature,
and $z_{ref}$ and $T_{ref}$ are the reference depth and temperature where
the reference viscosity is attained. The temperature dependence of the
viscosity is controlled by the activation energy while the pressure dependence
by the activation volume \citep[e.g.,][]{karato93,hirth13}.

At the temperature and pressure conditions of planetary mantles the viscosity
varies over orders of magnitude and is the parameter primarily controlling
the vigor of convection, the formation of a stagnant lid -- an immobile
layer at the top of the convecting mantle caused by the strong increase
of viscosity with decreasing temperature --, and the convection pattern in
the mantle. The latter is sensitive to the increase of the viscosity with
depth, and may result in a low degree convection pattern for a strong depth-dependent
viscosity or a sudden viscosity increase (i.e., a viscosity jump) in the
mid-mantle \citep{roberts06,keller09}. Such a convection pattern has been
previously proposed to explain the crustal thickness dichotomy
and focused volcanic activity in Tharsis, which is the largest volcano-tectonic
region on Mars
\citep[e.g.,][]{harder96,breuer98,zhong01,roberts06,keller09}.

Geodynamic thermal evolution models account for the decay of radioactive
elements with time and employ a cooling boundary condition at the CMB.
Using a 1-D energy balance the evolution of the CMB temperature is calculated
under the assumption of a constant core density and heat capacity
\citep{stevenson83,steinbach94}:
%
\begin{equation}\label{eq_core}
c_c\rho_cV_c\frac{dT_{CMB}}{dt} = -q_cA_c,
\end{equation}
where $c_{c}$, $\rho _{c}$, and $V_{c}$ are the core heat capacity, core density,
and core volume, respectively. $T_{CMB}$ and $q_{c}$ are the temperature and
the heat flow at the CMB, respectively, while $A_{c}$ is the CMB area.

Geodynamic models have been employed in previous studies to investigate
the effects of crustal thickness variations, as derived from gravity and
topography data, on the surface heat flow variations and the thermal state
of the lithosphere \citep{plesa16,plesa18b}. The crust in these studies
does not change with time, but varies spatially according to the chosen
crustal thickness model \cite[]{wieczorek04,wieczorek22}. Results show that
the crustal thickness variations and the crustal enrichment in HPEs control
the surface heat flow distribution at present day, the thickness of the
lithosphere and the lithospheric temperature variations. In addition to
global output quantities such as the average lithosphere thickness, average
surface and CMB heat flows and average mantle temperature, these models
can provide local values that can be compared to regional estimates. This
is essential to evaluate constraints provided by local measurements and
helps to put regional scale data in a global context.

\section{Crustal thickness estimates, partial melting, and the thermal state of the lithosphere}
\label{sec:Lith}

The thickness of the crust provides important constraints for later planetary
differentiation after core formation and for the overall thermal evolution
of the mantle. The crust that is built after the initial differentiation
of the planet and the crystallization of a potential magma ocean records
the magmatic activity through time, as it is formed by partial melting
of the mantle. During mantle melting, incompatible elements such as heat
producing elements (U$^{235,238}$, K$^{40}$, and Th$^{242}$) and volatiles
(e.g., H$_{2}$O and CO$_{2}$) are preferentially enriched in the melt.
The melt, due to its lower density compared to the surrounding mantle rises
to the surface, where it crystallizes and produces the crust. While volatiles,
such as H$_{2}$O and CO$_{2}$ are released in the atmosphere, the heat
producing elements (HPEs) remain stored in the crust that becomes more
enriched than the primitive mantle.

On Mars, the bulk of the crust has been built during the early history
\citep{greeley96,nimmo05} with an intense volcanic activity during Noachian.
Over time, volcanic activity declined and became more focused in Tharsis
and Elysium, the largest volcanic provinces on Mars. Young lava flows in
both Tharsis \citep{neukum04,hauber11} and Elysium \citep{vaucher09} indicate
that Mars has remained volcanicaly active over most of its history and
that partial melt production in the planet's interior may be ongoing. The crustal heat 
production rate that was derived from the surface abundances of thorium and potassium 
recorded by the gamma-ray spectrometer (GRS) on board Mars Odyssey \citep{taylor06} 
indicates a higher crustal enrichment factor compared to typical values for mid-ocean ridge 
basalts on Earth \citep{bvsp81}. The GRS data shows also a rather homogeneous distribution
with crustal thorium abundances between 0.2 and 1 ppm
\citep{taylor06b}, indicating much smaller variations than those observed on the Moon
\citep{lawrence00}. Based on this rather homogeneous distribution of thorium
and potassium at the surface of Mars and assuming that the surface abundance
of HPEs is representative for the entire crust, it was suggested that crustal
thickness variations have a stronger effect on the crustal heat flow variations \citep{hahn11}.

\begin{figure}[ht]%
\centering
\includegraphics[width=\textwidth]{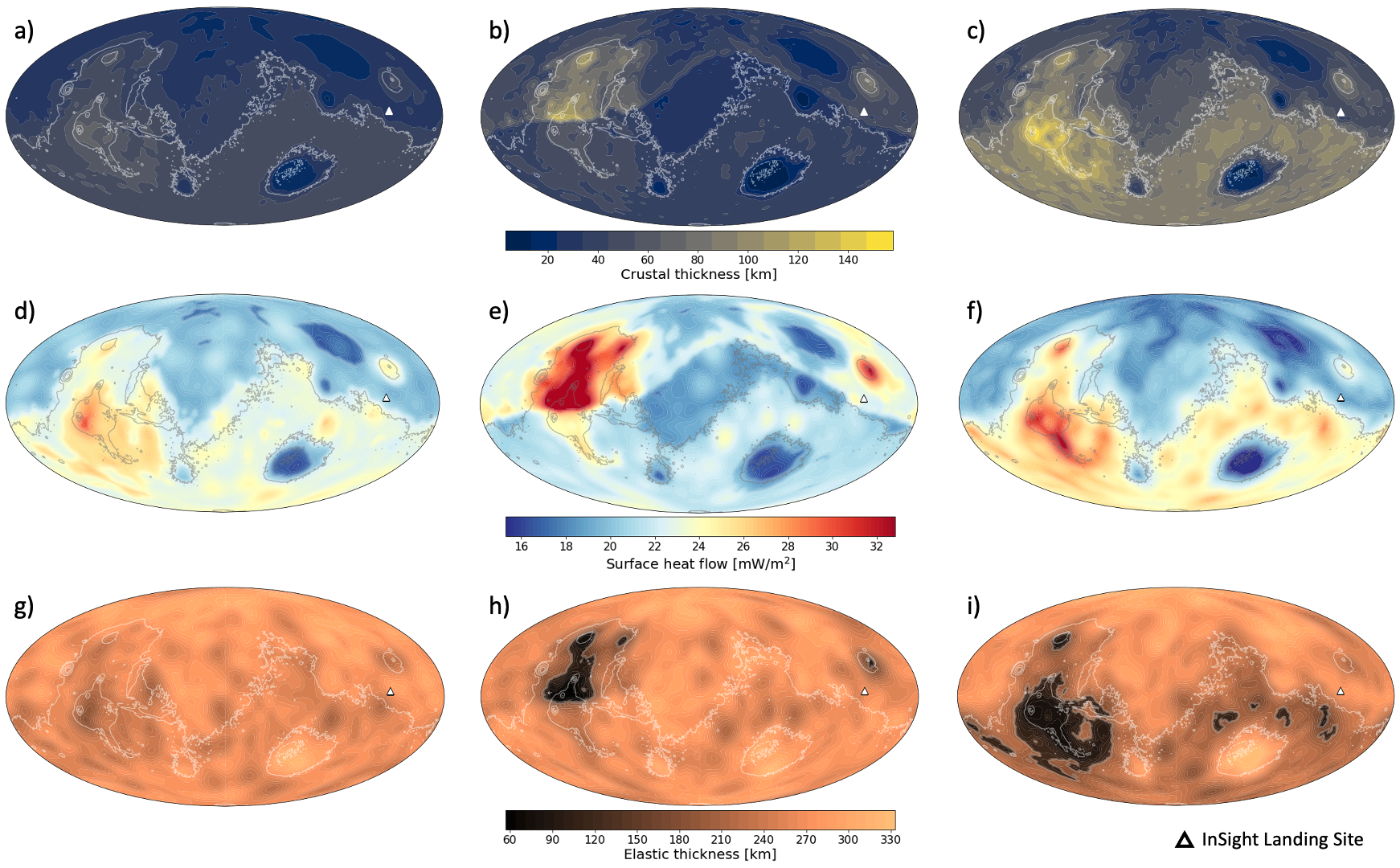}
\caption{Distribution of the crustal thickness (a, b, and c), associated
heat flow variations (d, e, and f) and elastic lithosphere thickness (g, h, and
i) at present day. Panels a, d, and g show the thin crustal thickness end-member
with an average crustal thickness of 40.6 km and a crustal density of 2550
kg\,m$^{-3}$. Panels b, e, and h present a crustal thickness model with and
average thickness of 43.1 km and a density difference between northern lowlands
(3000 kg\,m$^{-3}$) and southern highlands (2600 kg\,m$^{-3}$). In panels c, f,
and i an end-member model with an average crustal thickness of 71.4 km and a
crustal density of 3000 kg\,m$^{-3}$. The white and gray contour lines show the
0 km level of surface topography obtained from the Mars Orbiter Laser Altimeter
(MOLA) on board Mars Global Surveyor (MGS).}%
\label{fig:Cr_HF_Te}%
\end{figure}

Perhaps the most prominent geological feature on Mars is the crustal dichotomy.
The cause for the difference in elevation and crustal thickness between
the southern highlands and the northern lowlands is poorly known. Crustal
thickness models that can explain the gravity and topography data are nonunique.
While an anchor point given by a crustal thickness value at a known location
can help to constrain these models, one major assumption that remains is
the density of the crust and how it varies laterally \citep{wieczorek22}. When considering different crustal densities for the northern lowlands and southern highlands, the difference in crustal thickness across the dichotomy boundary can be small
(Fig.~\ref{fig:Cr_HF_Te}a), entirely absent (Fig.~\ref{fig:Cr_HF_Te}b),
or clearly visible (Fig.~\ref{fig:Cr_HF_Te}c).

Seismic data from the InSight mission have been used to determine the crustal
thickness at the InSight landing site \citep{knapmeyer-endrun21}. In the initial receiver function analyses, which are sensitive to the local crustal structure beneath the Insight lander
\citep{knapmeyer-endrun21}, two possible crustal models were premissible:
a two layer crust based on two strong seismic discontinuities below the surface at depths of about 8 and 20 km, or a three
layer crust that included a third weaker discontinuity recorded in the data at about 39 km depth. These seismic constraints can be used in combination with gravity and topography data to construct global crustal thickness models of the planet \citep{wieczorek04,wieczorek22}. For the two layer model, such modeling predicts an average crustal thickness somewhere
between 24 and 38 km \citep{knapmeyer-endrun21}, whereas for the three layer model the average thickness is predicted to be between 30 and 72 km \citep{wieczorek22}. We note that the
range of crustal thicknesses in \citet{wieczorek22} for the three layer model is slightly larger than the
range of 39 to 72 km presented in \citet{knapmeyer-endrun21}. This is because the initial study considered that the density of the crust was laterally homogeneous, whereas the latter study considered cases where the density of the crust could differ across the dichotomy boundary.

While none of the two crustal thickness models (two layer or three layer)
can be currently excluded based on the receiver function analysis, thermal
evolution models typically produce a thicker crust than the average crustal
thickness predicted by the two layer model \citep{knapmeyer-endrun21}.
Moreover, a recent study by \citet{kim21} that analyzed free-surface multiples
of the P-wave and combined these with receiver function analysis also favors
the three layer crust scenario. Therefore here we will discuss only models
using the three layer crust scenario.

In Fig.~\ref{fig:Cr_HF_Te} we show three crustal thickness models that
match the gravity and topography data, and that are anchored at the InSight
landing site by the three layer crustal thickness derived from the seismic observations.
The crustal thickness models represent end-members in terms of crustal
densities and crustal thicknesses. The thinnest crust uses a crustal thickness
at the InSight landing site of 31 km and the lowest considered crustal density
of 2550 kg\,m$^{-3}$ (Fig.~\ref{fig:Cr_HF_Te}a) leading to an average crustal
thickness of 40.6 km. An end-member model with a crustal thickness of 49
km at the InSight landing site and the highest crustal density value of
3000 kg\,m$^{-3}$ (Fig.~\ref{fig:Cr_HF_Te}c) leads to an average crustal
thickness of 71.4 km. In addition to the two end-member models, Fig.~\ref{fig:Cr_HF_Te}b shows another model that uses a crustal thickness at
InSight landing site of 47 km and different densities for the northern
lowlands (3000 kg\,m$^{-3}$) and southern highlands
(2600 kg\,m$^{-3}$). This model has an average crustal
thickness of 43.1 km. In this case, the dichotomy in crustal density largely
erases the crustal thickness dichotomy that is prominent in the other
two models in Fig.~\ref{fig:Cr_HF_Te}a and c.

Previously, crustal thickness models have been combined with geodynamic
models to investigate the effects of crustal thickness variations on the
surface heat flow variations and thermal state of the lithosphere
\citep{plesa16,plesa18b}. These models show that crustal thickness variations
are the main contributions to the heat flow variations and mantle plumes
have only a minor effect \citep{plesa16}. Here we discuss 3D thermal evolution
calculations similar to \cite{plesa16,plesa18b} that include updated crustal
thickness models, which have been constrained by InSight seismic data.
A list of the parameters used in the geodynamic simulations is shown in
Table~\ref{tab:Geodyn}, while details of the crustal thickness modeling
are discussed in \citet{wieczorek22} and crustal parameters are shown in
Table~\ref{tab:Crust}.

\begin{table}[htbp]
\caption{Parameters used in the geodynamic models in Fig.~\ref{fig:Cr_HF_Te}.}
\label{tab:Geodyn}
\begin{center}
\begin{tabular}{lll}
\hline
{\bf Symbol} & {\bf Description} & {\bf Value}\\
\hline
$D$ & Mantle thickness & $1550$\,km \\
$T_{ref}$ & Reference temperature & $1600$\,K \\
$p_{ref}$ & Reference pressure & $3$ $\times$ $10^{9}$\,Pa \\
$E$ & Activation energy & $3$ $\times$ $10^{5}$\,J\,mol$^{-1}$ \\
$V$ & Activation volume & $10$ $\times$ $10^{-6}$\,m$^3$\,mol$^{-1}$ \\
$T_{init}$ & Initial mantle temperature & $1800$\,K \\
$\Delta T$ & Initial temperature drop across the mantle & $2000$\,K \\
$\alpha$ & Reference thermal expansivity & $2.5$ $\times$ $10^{-5}$\,K$^{-1}$ \\
$\eta$ & Reference viscosity & $10^{21}$\,Pa\,s \\
$c_p$ & Mantle heat capacity & $1142$\,J\,kg$^{-1}$\,K$^{-1}$\\
$\rho$ & Mantle density & $3500$\,kg\,m$^{-3}$ \\
$c_c$ & Core heat capacity & $850$\,J\,kg$^{-1}$\,K$^{-1}$\\
$\rho_c$ & Core density & $6000$\,kg\,m$^{-3}$ \\
$g$ & Surface gravity acceleration & $3.72$\,m\,s$^{-2}$ \\
$k$ & Mantle thermal conductivity & $4$\,W\,m$^{-1}$\,K$^{-1}$ \\
$k_{cr}$ & Crust thermal conductivity & $3$\,W\,m$^{-1}$\,K$^{-1}$ \\
$\kappa$ & Mantle thermal diffusivity & $1$ $\times$ $10^{-6}$\,m$^2$\,s$^{-1}$ \\
$Q$ & Total initial radiogenic heating (mantle and crust) & $23.33$\,$\times$ $10^{-12}$ W\,kg$^{-1}$ \\
\hline
\end{tabular}
\end{center}
\end{table}

\begin{table}[htbp]
\caption{Parameters of the crustal thickness models shown in
Fig.~\ref{fig:Cr_HF_Te}. A detailed description of the crustal thickness models
is presented in \citet{wieczorek22}. $\rho _{N}$ and $\rho _{S}$ are the
densities of the northern lowlands and southern highlands, respectively, avg.
$d_{c}$ is the average crustal thickness, min. $d_{c}$ and max. $d_{c}$ are the
minimum and the maximum crustal thickness values, and $d_{c}^{InSight}$ is the
crustal thickness at InSight location.}
\label{tab:Crust}
\tabcolsep=3.5pt
\begin{tabular}{lcccccc}
\hline
{\bf Model} & {\bf $\rho_N$} & {\bf $\rho_S$} & {\bf avg. $d_c$} & {\bf min. $d_c$} & {\bf max. $d_c$} & {\bf $d_c^{InSight}$}\\
            &  [kg\,m$^{-3}$]    &     [kg\,m$^{-3}$] &    [km]             &      [km]           & [km]           & [km]\\
\hline
thin crust              & 2550 & 2550 & 40.6 & 11.7 & 72.4  & 31 \\
density dichotomy crust & 3000 & 2600 & 43.1 & 2.9  & 130.4 & 47 \\
thick crust             & 3000 & 3000 & 71.4 & 5.0  & 157.2 & 49 \\
\hline
\end{tabular}
\end{table}

In the absence of direct heat flow measurements, the elastic lithosphere
thickness at various times and locations can be used as a proxy for the
surface heat flow, as it can be linked to the thermal state of the lithosphere.
The elastic thickness characterizes the stiffness of the lithosphere in
response to loading and can be related to the mechanical thickness, given
a rheological model. The mechanical thickness is directly linked to the
thermal state of the lithosphere, since it can be identified with an isotherm
\citep{mcnutt84} and thus, it can be directly compared to lithospheric
temperatures from thermal evolution models (Fig.~\ref{fig:Cr_HF_Te}). This
comparison can be performed at various times during the evolution and at
different locations, depending on the time and location, for which elastic
thickness estimates are available.

The base of the mechanical lithosphere can be calculated following the
approach of \cite{grott08} and \cite{plesa16,plesa18} and assuming a bounding
stress of $\sigma _{B}$ of the order of $10^{7}$ Pa
\citep{grott10,burov95}, which gives the temperature associated with ductile
failure:
%
\begin{equation}
\label{eq_te}
T_{e} = \frac{E}{R}\left [\log\left (
\frac{\sigma ^{n}_{B}A}{\dot{\varepsilon}}\right )\right ]^{-1},
\end{equation}
where $E$, $A$, and $n$ are rheological parameters, $R$ is the gas constant,
and $\dot{\varepsilon}$ is the strain rate. A list of the rheological parameters
used to calculate the mechanical thickness is shown in Table~\ref{tab:TeParameters}.

\begin{table}[htbp]
\caption{Rheological parameters used for the calculation of the mechanical
lithosphere thickness \citep[][and references therein]{grott08}.}
\label{tab:TeParameters}
\tabcolsep=3pt
\begin{center}
\begin{tabular}{lll}
\hline
{\bf Symbol} & {\bf Description} & {\bf Value}\\
\hline
$E_{ol}$ & Activation energy, dry olivine dislocation creep & $5.4$ $\times$ $10^5$\,J\,mol$^{-1}$ \\
$E_{dia}$ & Activation energy, wet diabase dislocation creep & $2.76$ $\times$ $10^5$\,J\,mol$^{-1}$  \\
$A_{ol}$ & Prefactor, olivine dislocation creep & $2.4$ $\times$ $10^{-16}$\,Pa$^{-n}$\,s$^{-1}$ \\
$A_{dia}$ & Prefactor, diabase dislocation creep & $3.1$ $\times$ $10^{-20}$\,Pa$^{-n}$\,s$^{-1}$ \\
$n_{ol}$ & Stress exponent, dry olivine dislocation creep & $3.5$ \\
$n_{dia}$ & Stress exponent, wet diabase dislocation creep & $3.05$ \\
$\sigma_{B}$ & Bounding stress & $10^7$\,Pa \\
$\dot{\varepsilon}$ & Strain rate & $10^{-14}$\,s$^{-1}$ \\
\hline
\end{tabular}
\end{center}
\end{table}

The mechanical thickness represents an upper bound for the elastic lithosphere
thickness. However, it should be noted that the mechanical and elastic
thickness are similar for small curvatures and bending moments as it is
the case for the large geological features that are considered here
\citep{mcgovern04,belleguic05}. Thus in the following, we will use the
term ``elastic thickness''.

The elastic thickness of the mantle and crust can be determined using Eq.~(\ref{eq_te}) for their individual rheological parameters (Table~\ref{tab:TeParameters}). For the calculations presented here, we use parameters
for a dry olivine mantle and a wet diabase crust similar to
\citet{grott08}. These rheological parameters have been found to match
best the elastic thickness estimates available for the early history (Noachian
epoch) and present-day Mars \citep{grott10,breuer16,plesa18b}. If a layer
of incompetent crust separates the elastic cores of the mantle
$D_{m}$ and crust $D_{c}$, then the elastic thickness of the crust and
mantle system is significantly reduced and the effective elastic thickness
can be calculated as follows \citep{grott08,grott10}:
%
\begin{equation}
\label{eq_de}
D_{e} = (D_{m}^{3} + D_{c}^{3})^{\frac{1}{3}}.
\end{equation}

Otherwise, if the elastic thickness of the crust equals the crustal thickness,
then the effective elastic thickness is the sum of the two contributions.

On Mars, gravity and topography analysis, lithospheric flexure studies,
and estimates of the brittle to ductile transition indicate elastic lithosphere
thicknesses smaller than about 25 km during the Noachian epoch
\citep[][and references therein]{grott13}. These small values suggest a
warm lithosphere and/or a low mantle viscosity during the early martian
history \citep{grott13,thiriet18,plesa18b}. On the other hand, present-day
elastic thickness estimates that are available for the north and
south poles of Mars indicate a much thicker and colder lithosphere at these
two locations. This has been concluded based on the lack of downward deflection
with uncertainties of 100--200 m, beneath the north polar cap as seen by the
MARSIS and SHARAD radars \citep{phillips08}, while for the south polar cap
a maximum lithospheric flexure of 770 m has been found
\citep{broquet21}. Previous elastic lithosphere thickness estimates with
values larger than 300 km for the north pole \citep{phillips08} and larger
than 150 km for the south pole \citep{wieczorek08} have been reevaluated
in two recent studies \citep{broquet20,broquet21}. The latest estimates
indicate an elastic thickness between 330 km and 450 km for the north pole
of Mars \citep{broquet20} and a value larger than 150 km with a best fit
of 360 km for the south pole \citep{broquet21}.

The present-day elastic thickness estimates at the north and south poles
of Mars represent some of the strongest constraints for the thermal evolution
models \citep{plesa18b}. Successful models require that the elastic lithosphere
thickness values at these two locations are compatible with the present-day estimates. 
The elastic thickness is anti-correlated to the crustal thickness and surface heat flow (Fig.~\ref{fig:Cr_HF_Te}). Regions of thick crust typically associated 
with the southern hemisphere and in particular with volcanic centers show an elevated 
heat flow and a thin elastic thickness compared to the northern hemisphere and in 
particular within impact basins. This is due to the fact that a thicker crust has
a higher amount of HPEs and a stronger blanketing effect than a thinner
crust. The crustal blanketing effect is produced by the lower crustal conductivity
compared to that of the mantle. This leads to higher subsurface temperatures
in regions covered by a thick crust compared to areas with a thin crust.

The magnitude of surface heat flow and elastic thickness variations depends
on the magnitude of crustal thickness variations. In the following, we
discuss the effects of crustal thickness variations for the present-day
surface heat flow and elastic thickness pattern taking as examples three
geodynamic simulations that use the three different crustal thickness models
presented in Fig.~\ref{fig:Cr_HF_Te}. The geodynamic models use the bulk
heat production rate of \citet{taylor13} and assume that the mantle contains
about 43\% of the bulk heat production rate, a value that lies in the range
suggested by \citet{knapmeyer-endrun21} to produce localized melting regions
in the interior at present-day. A crustal thickness dichotomy leads to
a dichotomy in surface heat flow and elastic thickness. The smallest surface
heat flow and elastic thickness variations are obtained for the thinnest
crust scenario (Fig.~\ref{fig:Cr_HF_Te}d,~g), where crustal thickness variations
are more than a factor two smaller compared to the thickest crust scenario.
The crustal thickness variations for the case where the crustal density differs across the dichotomy boundary are about
25 km smaller than for the thickest crust scenario. However surface heat
flow and elastic thickness variations are more pronounced in this case
due to the difference in the crustal density and therefore the amount
of crustal HPEs between the southern and northern hemispheres. Due to the lower
crustal density of the southern compared to the northern hemisphere, the
volumetric heat production in the northern crust is higher than in the
southern crust leading to a warmer crust on the northern compared to the
southern hemisphere. This is reflected also by the higher surface heat
flow in the northern part of the Tharsis region compared to the southern
part.

\begin{table}[htbp]
\caption{Results obtained of the crustal thickness models shown in
Fig.~\ref{fig:Cr_HF_Te}. All values represent present-day values. $F_{s}$\,[$
\min , \max $] is the average surface heat flux with minimum and maximum values.
$T_{e}$\,[$\min , \max $] is the average elastic thickness with minimum and
maximum values calculated assuming a strain rate
$\dot{\varepsilon}=10^{-14}$\,s$^{-1}$. ${F_{s}}^{InSight}$ is the surface heat
flux at InSight location. ${T_{e}}^{NP}$ is the elastic lithosphere thickness
averaged below the north pole ice cap (i.e., within 10$^{\circ}$ from the north
pole), while ${T_{e}}^{SP}$ is the elastic lithosphere thickness averaged below
the south pole ice cap (i.e., within 5$^{\circ}$ from the south pole).
$T_{CMB}$ is the core-mantle boundary temperature and $F_{CMB}$ is the
core-mantle boundary heat flux.}
\label{tab:TeResults}
\tabcolsep=2pt
\begin{center}
\begin{tabular}{l|ccc}
\hline
\bf Output & \bf thin crust & \bf density dichotomy crust & \bf thick crust \\
\hline
$F_{s}$\,($\min, \max$) [mW\,m$^{-2}$] & 22.1\,(16.3,\,30.0) & 22.2\,(16.0,\,38.9) & 22.3\,(14.4,\,33.1) \\
$T_e$\,($\min, \max$) [km]             & 267\,(\hspace*{0.15cm} 185,\, 326) & 261\,(\hspace*{0.15cm}  61,\, 328) & 234\,(\hspace*{0.15cm}  70,\, 348) \\
${F_s}^{InSight}$ [mW\,m$^{-2}$]       & 20.1 & 22.6 & 19.1 \\
${T_e}^{NP}$ [km]                      & 284  & 288  & 304 \\
${T_e}^{SP}$ [km]                      & 264  & 280  & 236 \\
$T_{CMB}$ [K]                          & 2094.2 & 2092.7 & 2086.1 \\
$F_{CMB}$ [mW\,m$^{-2}$]               & 2.3 & 2.4 & 2.4 \\
\hline
\end{tabular}
\end{center}
\end{table}

The elastic thickness is thickest in areas of thin crust where the interior
cools more efficiently. These areas are typically impact basins, with the
Hellas impact basin usually recording the highest elastic thickness values.
The thinnest elastic thickness is obtained in areas covered by a thick
crust, where the presence of an incompetent crustal layer (i.e., a weak
crustal layer formed by high crustal temperatures) may decouple the elastic
cores of the mantle and the crust, thus further reducing the elastic thickness.
This has been suggested to exist at present day in the Tharsis area around
Arsia Mons, where the crust is thickest \citep{grott10}. Indeed such an
incompetent crustal layer is present in the crustal density dichotomy model
and the thickest crust scenario. While in the former this is located in
Tharsis and in a small area in Elysium, for the latter the incompetent
crustal layer is present in the Tharsis area and in smaller locations in
the southern hemisphere due to the overall thicker crust in this scenario.
As discussed in previous studies, the strongest constraint is given by
the elastic thickness at the north pole. While at the south pole, all models
present an elastic thickness greater than 150 km being compatible with
the latest estimate \citep{broquet21}, only the thickest crust scenario
presents an elastic thickness larger than 300 km (i.e., 304 km) at the
north pole (Table~\ref{tab:TeResults}), a value that is still lower than
the recent estimate of \citet{broquet20}. A higher elastic thickness may
be obtained if the mantle is more depleted in HPEs than assumed in these
models. However, a lower heat production in the mantle might lead to scenarios
in which partial melt production stops earlier than suggested by the geological
record in Tharsis and Elysium \citep{neukum04,vaucher09,hauber11}. Another
solution to explain the discrepancy between the elastic thickness estimates
and the values obtained from geodynamic models would require that the
load produced by the polar cap is not yet at elastic equilibrium as discussed
by \citet{broquet20}. This would lead to lower elastic thickness estimates,
since in this case the observed deflection would be the sum of a downward
deflection caused by the viscous relaxation and an upward deflection caused
by some form of postglacial rebound. Whether the north pole is at elastic
equilibrium strongly depends on the viscosity of the lithosphere and mantle
that is linked to the parameters of the geodynamic model and would require
the computation of an individual elastic thickness estimate for each thermal
evolution model. Nevertheless, future work needs to address this aspect,
since this may significantly affect the number of admissible models that
can explain the elastic lithosphere thickness at the north pole of Mars.

In addition to the surface heat flow and elastic lithosphere, the thickness
of the crust and its variations can affect the amount and distribution
of partial melt that may still be produced in the interior of Mars today.
Due to the pronounced crustal blanketing effect and higher amount of crustal
HPEs, regions covered by a thick crust can be kept warm and their temperatures
can exceed the melting temperature and produce melt up to recent times.
Whether melt can still be produced in the martian mantle at present day
primarily depends on the amount of mantle HPEs.

The study by \citet{knapmeyer-endrun21} showed that only a limited range
of crustal enrichment, i.e., containing between 55--70\% of the total bulk
of HPEs would lead to localized partial melt production at present day
in the interior of Mars. A strong crustal enrichment containing more than
70\% of the bulk amount of HPEs would lead to a mantle that is too cold
to produce melt at present day. On the other hand a mantle containing more
than 45\% of the bulk amount of bulk amount of HPEs would be too warm and
lead to wide-spread melting at present day. For a crust with an average
thickness at the upper end of values obtained from InSight's seismic data
this indicates a crustal enrichment in thorium and potassium similar to
the surface abundance as measured by the gamma-ray instrument on board
Mars Odyssey \citep{hahn11,taylor06}. A thinner crust, on the other hand,
requires an enriched component in the subsurface in order to avoid wide-spread
melting in the interior of Mars at present day
\citep{knapmeyer-endrun21}.

\begin{figure}[htbp]%
\centering
\includegraphics[width=\textwidth]{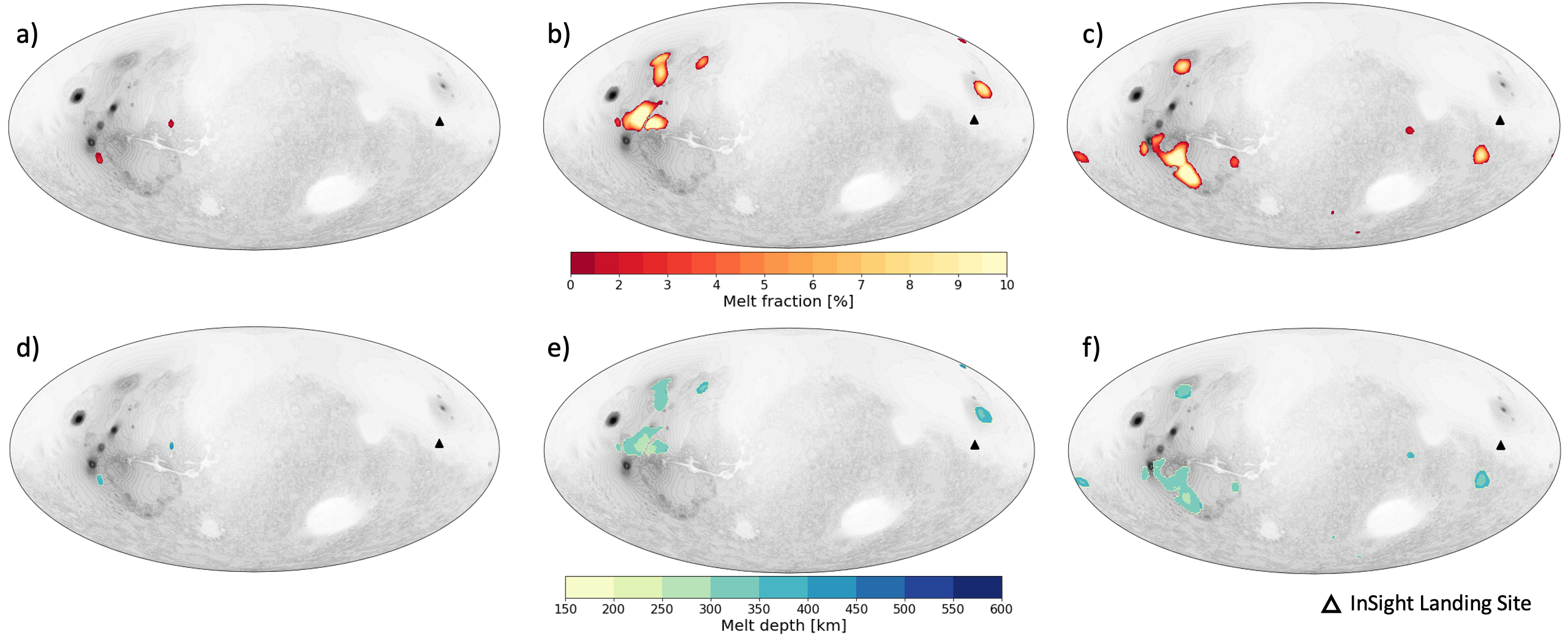}
\caption{Distribution of partial melt zones at present day in the mantle for the
models presented in Fig.~\ref{fig:Cr_HF_Te}. Panel a, b, and c show the melt
fraction and panels d, e, and f show the corresponding melt depth. The thin
crust end-member is shown in panels a and~d, the density dichotomy crust in
panels b and e, and the thick crust end-member in panels c and f.}%
\label{fig:Melt}%
\end{figure}

In Fig.~\ref{fig:Melt} we show the present-day melt fraction and the depth
of the melt zone for the three models presented in Fig.~\ref{fig:Cr_HF_Te}. The melting temperature is taken from
\citet{ruedas17}, who updated the solidus parametrization of
\cite{ruedas13} to include more recent melting experiments from
\citet{collinet15} and \cite{matsukage13}. Furthermore, \citet{ruedas17} include
a correction to account for the effects of Na, K, and Ca as suggested by
\citet{kiefer15} that leads to about 35 K lower solidus for the primitive
martian mantle compared to the terrestrial mantle. Additionally, the solidus
used in each model in Fig.~\ref{fig:Melt} considers the effect of mantle
depletion due to crust formation. The solidus is increased linearly with
the degree of depletion that each model experienced according to the crustal
volume.

The least melt is produced at depths larger than 350 km in the thin crust
model (Fig.~\ref{fig:Melt}a and d) that also has the smallest crustal thickness
variations. The crustal blanketing effect is more pronounced for the other
two models using a crustal density dichotomy and a higher density crust,
respectively, due to a locally thicker crust in these models. This results
in a larger amount of partial melt (higher melt fractions and more melt
regions) and shallower melt zones compared to the thin crust model.

In general, geodynamic models have difficulties to produce partial melt zones at present-day
beneath the Elysium volcanic province, even though mantle plumes are present
there. Melting takes place mostly in Tharsis and underneath the southern
hemisphere, as these areas are typically covered by a thicker crust. Since
the Elysium province lies in the northern lowlands, it is difficult to
focus mantle plumes underneath it that are hot enough to produce melt at
present day. Interestingly, the crustal density dichotomy model shows melting
zones focused in Tharsis and Elysium. The melting zone in Elysium is likely
due to a thicker crust in this region compared to the southern hemisphere
and due to the reduced differences in crustal thickness between the northern
and southern hemispheres caused by using a lower crustal density in the
south compared to the north. We note, however, that the density difference
between the southern and the northern crust is quite extreme in this model
(i.e., 400 kg m$^{-3}$), and whether a crustal thickness model with smaller
density variations between north and south can produce a partial melting
zone at present day beneath Elysium needs to be tested by future studies.

\section{Mantle dynamics, seismic velocities variations, and tidal dissipation}
\label{sec:MantleDyn}

\begin{figure}[htbp]%
\centering
\includegraphics[width=\textwidth]{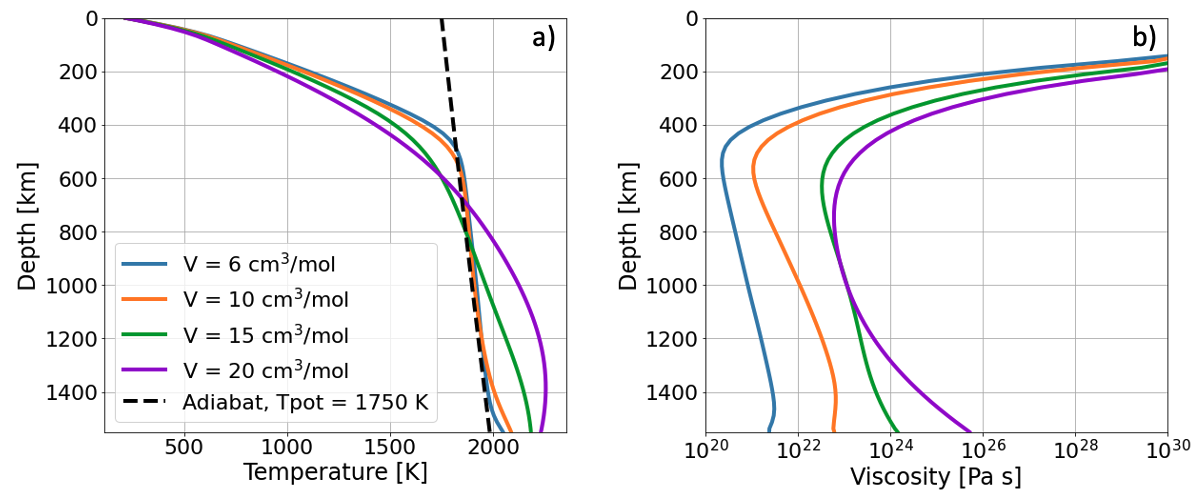}
\caption{Effect of the pressure dependence of the viscosity, which is given by
the activation volume, on the temperature (a) and viscosity (b) profiles at
present day. The values for the activation volume are taken from
\citet{hirth13}.}%
\label{fig:TVisc}%
\end{figure}

The thermal state of the mantle, the thermal lithosphere
thickness, as well as the location of hot mantle plumes and cold downwellings
can affect the variation of seismic velocities. Typically, the thickness
of the thermal lithosphere is the sum of the stagnant lid thickness and
of the thermal boundary layer where convective instabilities initiate.
Below the thermal lithosphere, the mantle temperature usually follows an
adiabatic profile. However, deviations from an adiabatic temperature profile
may occur if convection is sluggish due to a high pressure-dependence of
the viscosity or due to strong cooling of the interior, in which case the
average temperature profile lies between an adiabatic and a conductive
profile (Fig.~\ref{fig:TVisc}). Parametrized thermal evolution 
models typically use either an adiabatic mantle temperature profile or a conductive 
one, if convection stops (i.e., the Rayleigh number drops below a critical value). 
The 2D/3D geodynamic models, on the other hand, self-consistently calculate the thermal 
profile, and in these models, depending on the rheological parameters, the thermal 
profile in the mantle may lie between an adiabatic and a conductive profile.

\begin{figure}[htbp]%
\centering
\includegraphics[width=\textwidth]{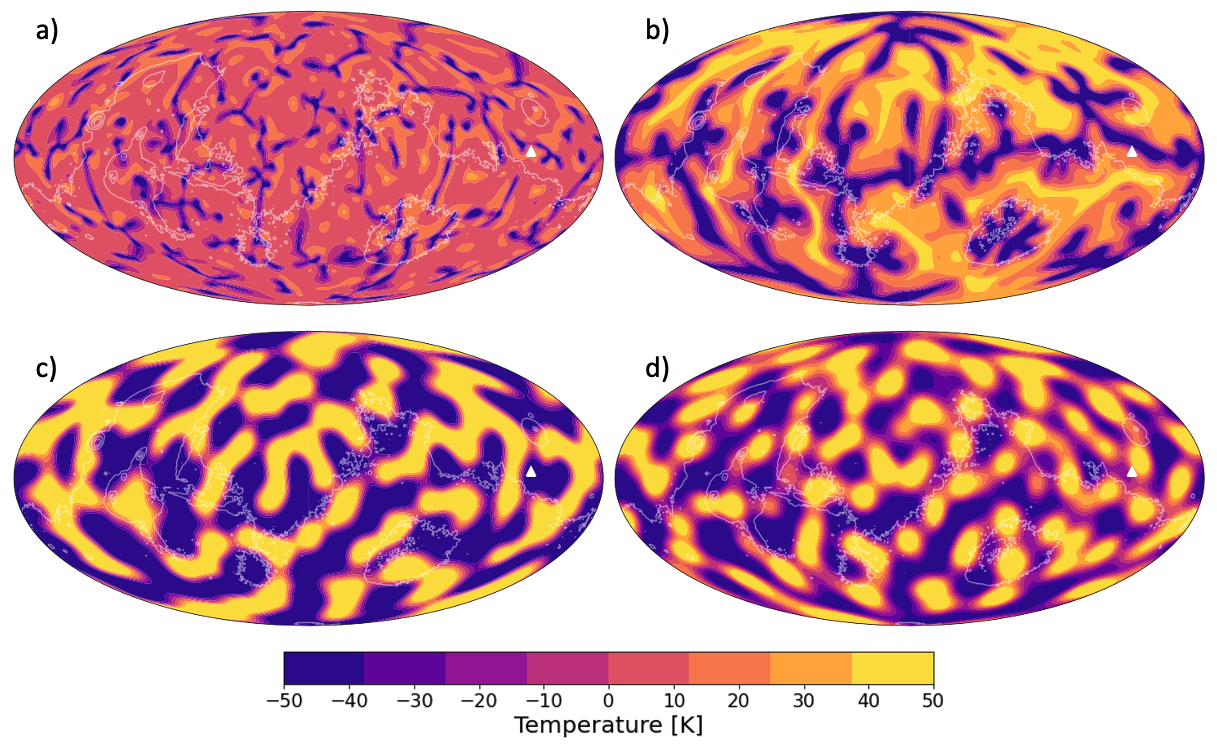}
\caption{Temperature variations at mid-mantle depth (775 km depth) corresponding
to the temperature profiles shown in Fig.~\ref{fig:TVisc}. The models use an
activation volume of 6 cm$^{3}$\,mol$^{-1}$ (panel a), 10 cm$^{3}$\,mol$^{-1}$
(panel b), 15 cm$^{3}$\,mol$^{-1}$ (panel c), and 20 cm$^{3}$\,mol$^{-1}$ (panel
d). The color scale has been clipped to show the locations of mantle plumes
(bright colors) and downwellings (dark colors). For orientation the 0 km level
of the surface topography is indicated by white contour lines.}%
\label{fig:TViscPattern}%
\end{figure}

The thermal state of the interior is to first order affected by the mantle
viscosity. However, rheological parameters such as the activation energy
and activation volume that are determined by laboratory deformation experiments
have large uncertainties. While the activation energy of olivine aggregates
was measured to lie at about $375\pm 75$~kJ\,mol$^{-1}$ for diffusion and
$520\pm 40$~kJ\,mol$^{-1}$ for dislocation creep \citep{hirth13}, the activation
volume is one of the most poorly constrained parameters with values between
0 and 20 cm$^{3}$\,mol$^{-1}$ \citep{hirth13}. These values will substantially
affect the mantle temperature profile (Fig.~\ref{fig:TVisc}) and also the
mantle convection pattern. An activation volume of 6 cm$^{3}$\,mol$^{-1}$
leads to a larger number of plumes and downwellings compared to an activation
volume of 10 cm$^{3}$\,mol$^{-1}$, as illustrated in the temperature maps
at mid-mantle depth (Fig.~\ref{fig:TViscPattern}a and b). However, further
increasing the activation volume would lead to scenarios, in which convection
in the lower part of the mantle is weak or even absent. This reduces the
thickness of the convective layer and increases the wavelength of the convection
pattern (Fig.~\ref{fig:TViscPattern}c and d).

The viscosity could also be affected by the presence of water in the mantle.
Water concentrations in excess of 100 ppm can reduce the viscosity by about
two orders of magnitude \citep{karato93} and can significantly affect mantle
cooling. On Mars, recent petrological analyses of martian meteorites suggest
a bulk water content of 137 ppm, with crustal abundances of 1410 ppm and
mantle water contents between 14 and 72 ppm \citep{mcCubbin16}. A wet mantle
rheology during most of the thermal history would not be able to reproduce
strong mantle plumes in recent times \citep{plesa18b} as required by the
petrological evidence for local mantle temperatures
\citep{filiberto15,kiefer16}. In addition, a dry rheology was favored by
models that coupled the thermal and orbital evolution of Mars and its moon
Phobos, in order to reproduce the orbital evolution of Mars' closest satellite
\citep{samuel19}. Thus, according to thermal evolution models, water in
the martian mantle was most likely lost during the earliest planetary evolution
and most of the thermal history was characterized by a dry mantle rheology.
We note, however, that geochemical reservoirs may complicate this interpretation,
as they could trap water in isolated regions inside the mantle and lithosphere
\citep{breuer16}.

\begin{figure}[htbp]%
\centering
\includegraphics[width=\textwidth]{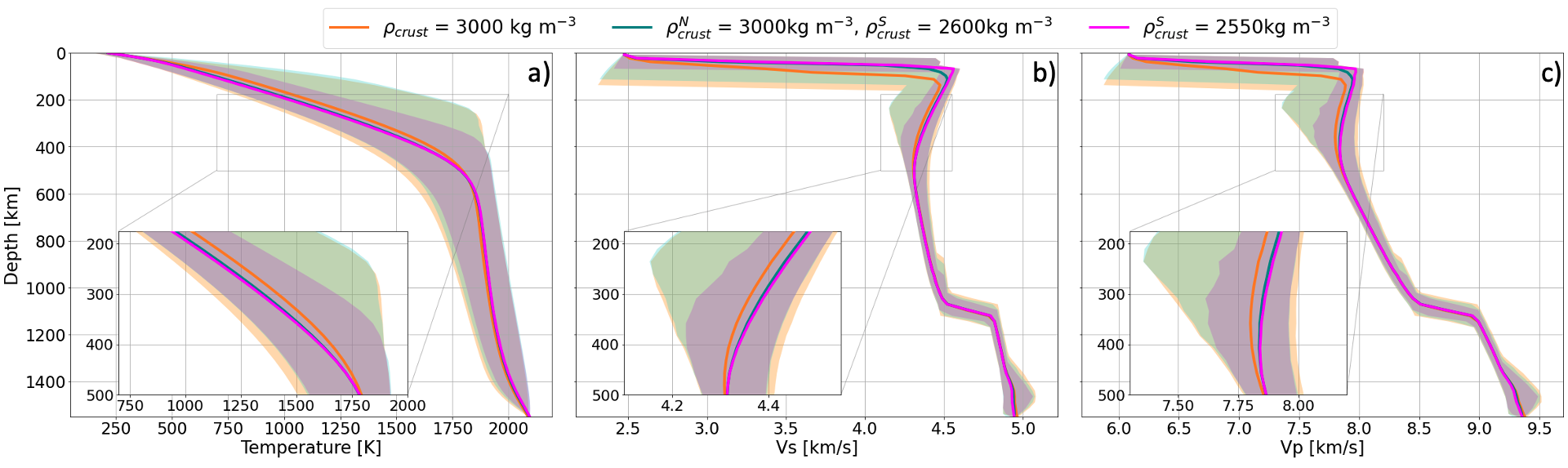}
\caption{Profiles of the temperatures (panel a) as well as the shear wave
velocities (panel b) and compressional wave velocities (panel c) for the models
presented in Fig.~\ref{fig:Cr_HF_Te}. The TAY13 \citep{taylor13} was assumed for
the seismic velocities calculation. Full lines indicate the average profiles
throughout the mantle, while the shaded regions show the corresponding
variations.}%
\label{fig:SeisVels}%
\end{figure}

The present-day thermal state of the interior is the result of billion
years of thermal evolution. Mantle plumes and cold downwellings may be
present in the interior of Mars at present day and would lead to temperature
variations in the interior and to variations in the thermal lithosphere
thickness. These effects can only be investigated by using 2D and 3D geodynamic
models. Previous models showed that the thermal lithosphere can be substantially
thinner at the location of mantle plumes \citep{kiefer09}, as their higher
temperature decreases locally the viscosity and allows the silicate material
to flow (i.e., to convect) at shallower depths. Conversely, the thermal
lithosphere is thicker above cold mantle downwellings. Additionally, the
variations of the thermal lithosphere thickness are affected by the crustal thickness variations, as the latter has a higher amount of HPEs and a lower
conductivity compared to the mantle. This leads to a higher lithospheric
temperature and thinner lithospheric thickness beneath areas covered by
a thick crust (e.g., beneath volcanic provinces) compared to regions of
thin crust (e.g., impact basins). For models that include crustal thickness
variations, the largest variations in temperature are observed in the lithosphere.
We also note that the largest temperature variations are obtained for models
with a thick crust that exhibits larger crustal thickness variations. In
Fig.~\ref{fig:SeisVels}a we show average temperature profiles and corresponding
temperature variations at present day for the three thermal evolution models
presented in Fig.~\ref{fig:Cr_HF_Te}. The model with a crustal density
dichotomy between the northern and southern hemisphere and the model with
a crustal density of 3000 kg\,m$^{-3}$ have crustal thickness variations
of 127.2~km and 152.3~km, respectively, that lead to larger lithospheric
temperature variations than the crustal thickness model with a crustal
density of 2550 kg\,m$^{-3}$, which has a difference of only 60.8 km between
the minimum and maximum crustal thickness values.

\begin{figure}[htbp]%
\centering
\includegraphics[width=\textwidth]{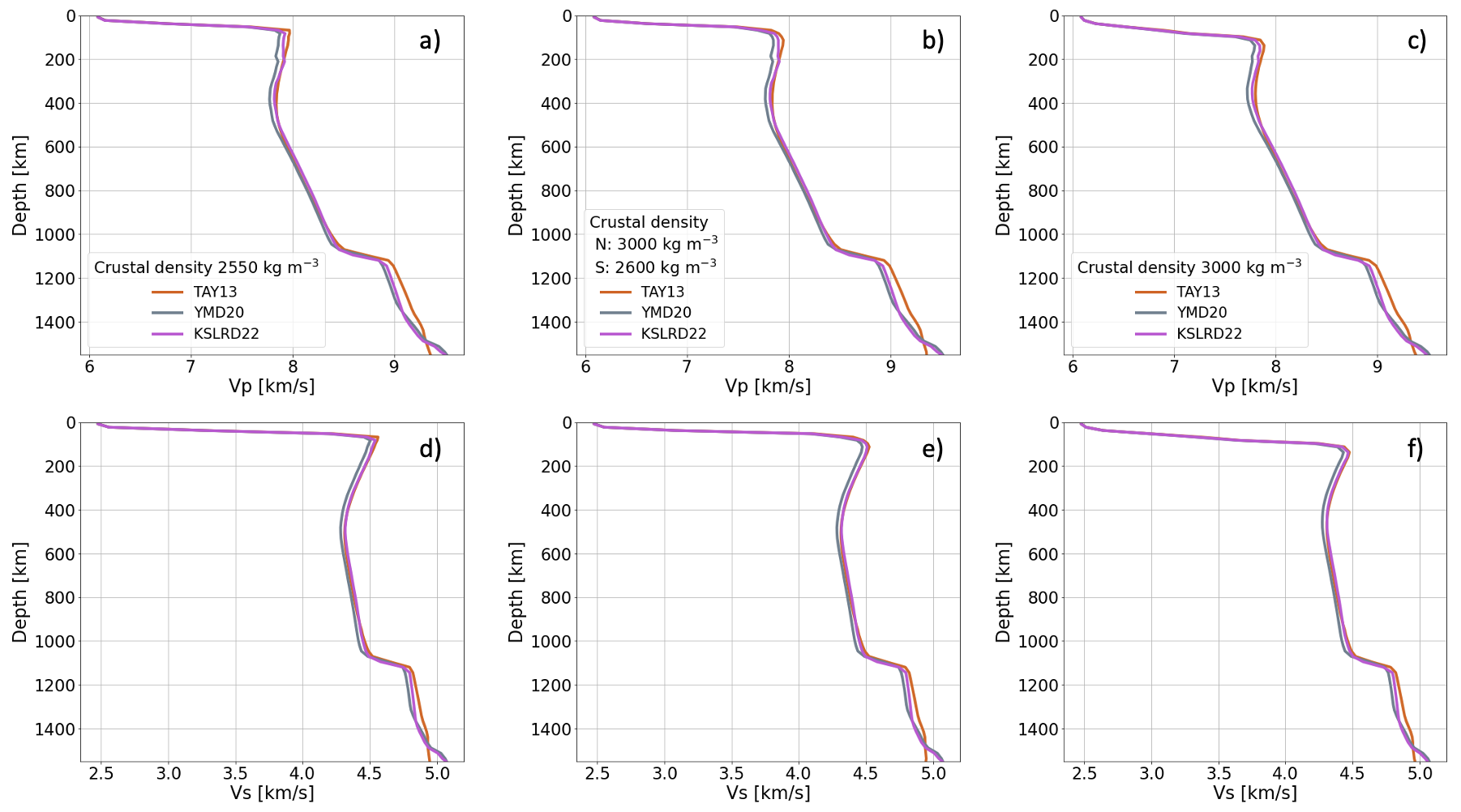}
\caption{Effects of mantle composition on the compressional wave velocities
(panels a, b, and c) and shear wave velocities (panels d, e, and f) for three
different mantle compositions: TAY13 \citep{taylor13}, YMD20
\citep{yoshizaki20}, and KSLRD22 \citep{khan22}. Panels a and d show the thin
crust end-member case, panels b and e present the results for the density
dichotomy crust, and panels c and f show the thick crust end-member case.}%
\label{fig:SeisVelsCompo}%
\end{figure}

On Mars, the average thermal lithosphere thickness has been estimated based
on the evaluation of seismic events recorded by InSight. Using direct and
surface-reflected body wave phases, a thermal lithosphere thickness of
400--600~km is required to explain the differential travel times obtained for
seismic events at epicentral distance between 25$^{\circ}$ and 75$^{\circ}$
from InSight location and with moment magnitudes between 3 and 4
\citep{khan21}. InSight's estimate of the average thermal lithosphere
thickness of Mars is thicker than the thermal lithosphere on the Earth and
suggests that Mars has significantly cooled during its thermal history. This
thick lithosphere and its thermal gradient inferred from InSight data control
the formation of low-velocity zones in the interior of Mars that have been
proposed in previous studies \citep{mocquet00,zheng15}. Additionally, a
recent study by \citet{plesa21} showed that 3D thermal evolution models with a crust containing
less than 20\% of the bulk amount of HPEs would lead to a hot interior and a
thin lithosphere. These models are incompatible with InSight observations, as
they would lead to S-wave shadow zones for high-quality events in the
Cerberus Fossae region, for which clear P- and S-waves arrivals were
recorded.

Temperature variations affect the seismic velocities with the strongest
velocities variation being present in the lithosphere and at the depth
of the olivine to wadsleiyte phase-transition (Fig.~\ref{fig:SeisVels}b and c). We note that the effects of composition are
minor compared to the effect of temperature variations in the lithosphere.
In Fig.~\ref{fig:SeisVelsCompo}, we show the differences between the average
seismic velocities profiles for the three models presented in Fig.~\ref{fig:SeisVels}. We tested three of the most recent compositions that
have been proposed for Mars: the TAY13 composition \citep{taylor13}, the
YMD20 composition \citep{yoshizaki20}, and the KSLRD22 composition
\citep{khan22}. While the seismic velocities have been computed using these
compositions, the bulk amount of HPEs that was used in all thermal evolution
models was taken from TAY13. Other HPEs models with a higher abundance
of radiogenic elements such as the model by \citet{yoshizaki20} would require
a higher crustal enrichment in HPEs and a similar amount of mantle HPEs
as the models shown here in order to avoid wide-spread melting in the mantle
at present day. Thus, even for other HPEs models the mantle temperature
would be similar to the profiles shown in Fig.~\ref{fig:SeisVels}.

For all three compositional models tested here, the seismic velocities
are nearly identical in the upper mantle and show slight differences in
the lower mantle, with minimally lower seismic velocities values for the
YMD20 and KSLRD22 compositions mainly due to the lower FeO content of these
models (14.7 $\pm $ 1.0 wt\% for YMD20 and 13.7 $\pm $ 0.4 wt\% for KSLRD22
compared to 18.1 $\pm $ 2.2 wt\% for TAY13).

\begin{figure}[htbp]%
\centering
\includegraphics[width=\textwidth]{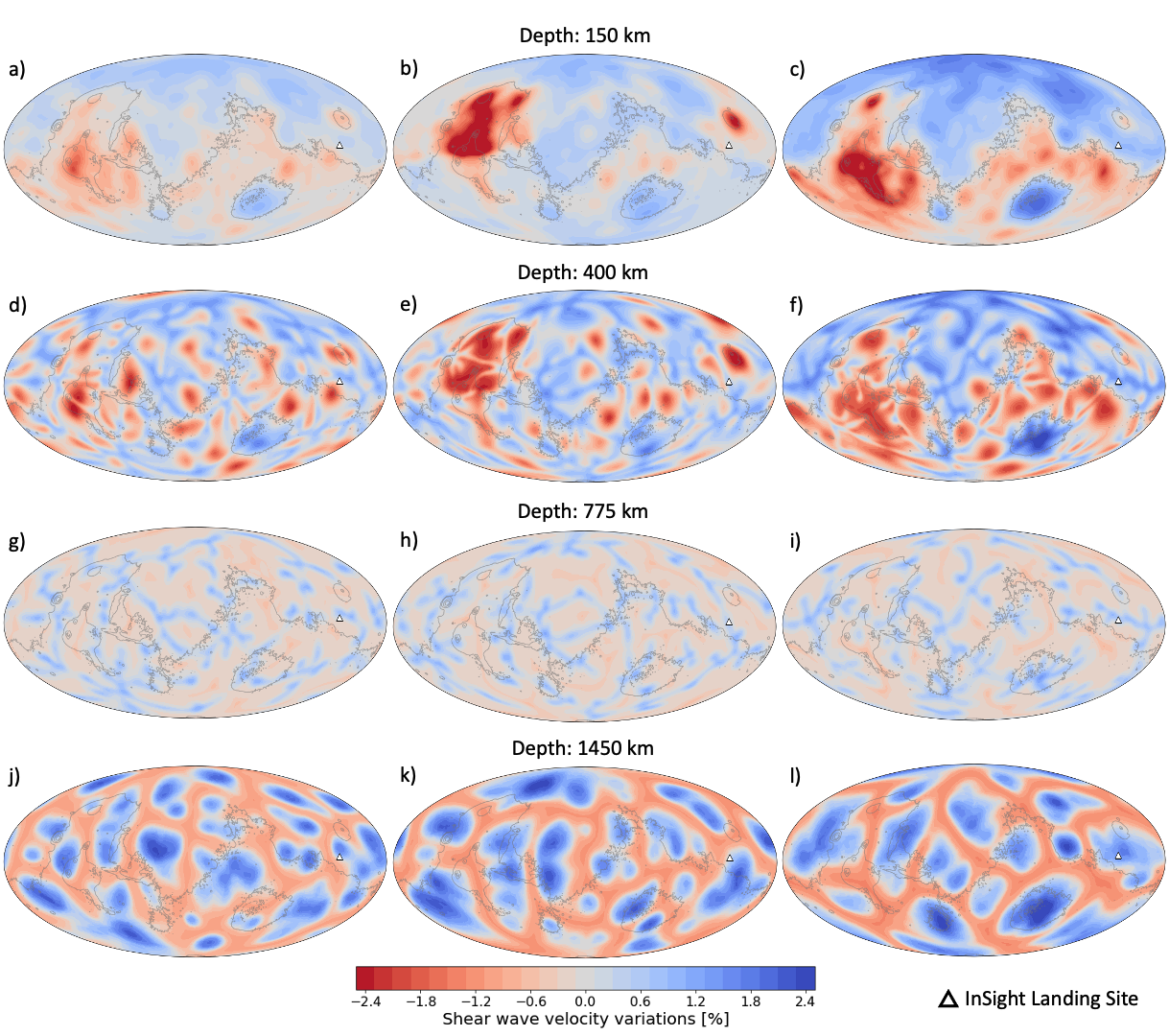}
\caption{Seismic velocities variations at different depths throughout the mantle
calculated for TAY13 composition. Left column (panels a, d, g and j) shows the
thin crust end-member case. Middle column (panels b, e, h, and k) presents the
density dichotomy crust model, and the right column (panels c, f, i, and l) shows
the thick crust end-member case. For orientation, the gray contour lines show
the 0 km level of surface topography.}%
\label{fig:SeisVelsMaps}%
\end{figure}

The temperature and hence the seismic velocities in the lithosphere follow
the crustal thickness variations. The crustal thickness pattern controls
their variations down to a depth of 400 km or even deeper in particular
for models with a thick crust (Fig.~\ref{fig:SeisVelsMaps}). In Fig.~\ref{fig:SeisVelsMaps}, the thin crust model (average crustal thickness
of 40.6 km, left column) shows a pattern of the S-wave velocity variations
that closely follows the crustal thickness pattern at 150 km depth. Lower-than-average
seismic velocities are observed below the southern hemisphere
and are caused by the warmer temperatures due to a thicker crust compared
to the northern hemisphere. Conversely, the areas covered by a thin crust,
i.e., the northern hemisphere and large impact basins such as Hellas, show
seismic velocities higher than the average value, due to the more efficient
cooling and hence colder temperatures than those beneath the southern hemisphere.
This seismic velocities pattern is no longer visible at 400 km depth. For
the thick crust model (average crustal thickness of 71.4 km, right column),
however, a dichotomy in the S-wave velocity variations is still visible
at 400 km depth. Since all models in Fig.~\ref{fig:SeisVelsMaps} use the
same bulk heat production and the same amount of HPEs in the mantle, the
difference is caused by the stronger crustal thickness variations and the
more pronounced blanketing effect due to the low crustal conductivity in
the thick crust scenario compared to the thin crust case. In the crustal
density dichotomy case (middle column), the crustal thickness pattern is
more complex than in the previous two models, but the Tharsis region is
clearly distinguishable on the map of S-wave velocity variations at 400
km depth. Seismic velocities variations are small and about 1\% in the
convecting mantle. Larger variations are observed again closer to the CMB,
where negative seismic velocity gradients may be locally present due to
the stability of larger proportions of garnet and ferropericlase at the
expense of ringwoodite \citep{plesa21}. We note however that this depends
on the chosen mineralogical model and mostly appears for the Taylor-composition
\citep{taylor13}, but is absent for the Yoshizaki- and Khan-compositions
\citep{yoshizaki20,khan22}.

\begin{figure}[htbp]%
\centering
\includegraphics[width=\textwidth]{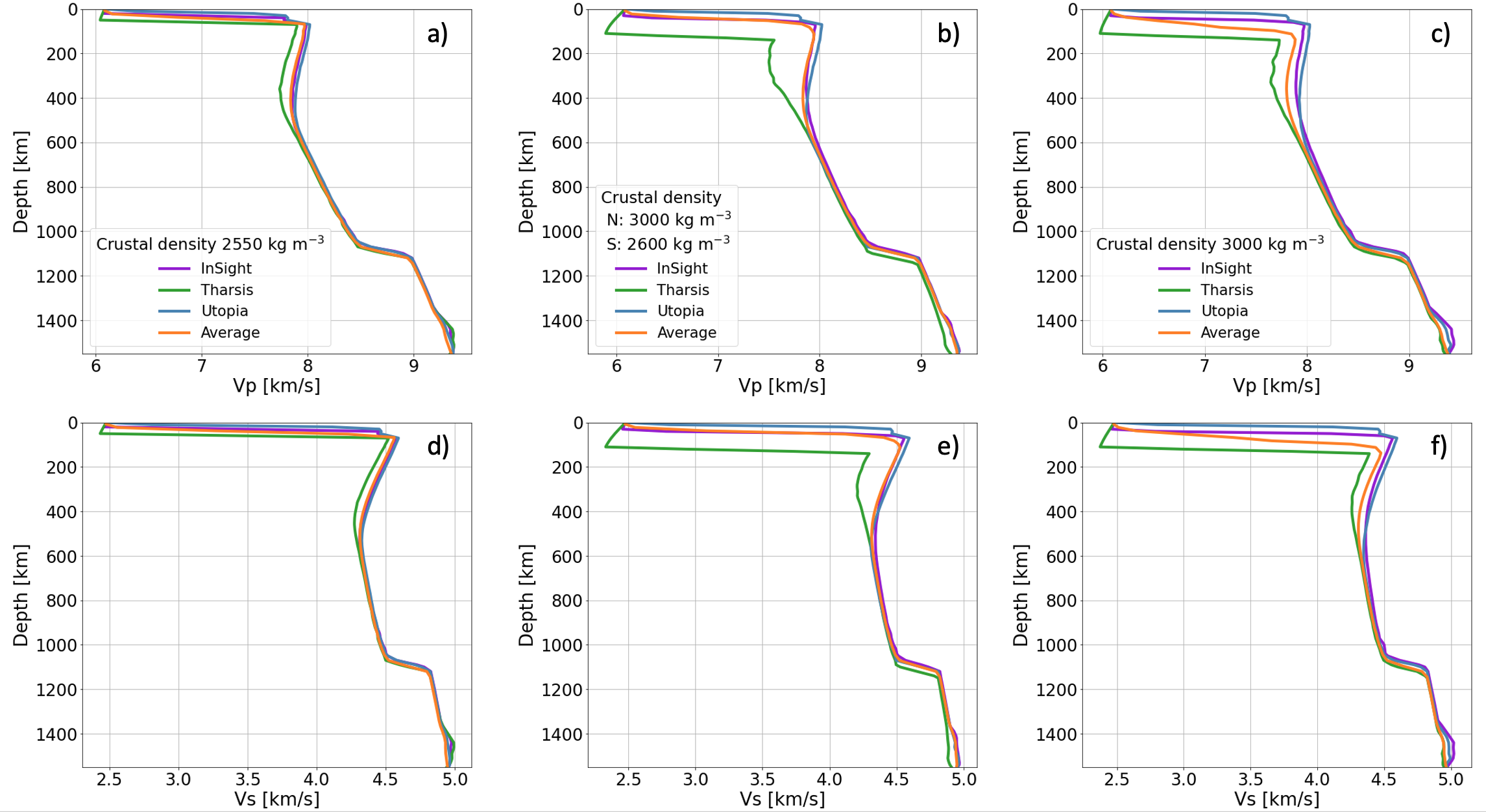}
\caption{Seismic velocities profiles at selected locations compared to the
average profiles. InSight profile is drawn at 136$^{\circ}$ longitude and
4$^{\circ}$ latitude. Tharsis profile is taken at $-$115$^{\circ}$ longitude and
0$^{\circ}$ latitude, while the Utopia profile was selected at 115$^{\circ}$
longitude and 45$^{\circ}$ latitude. Panels a and d show the seismic velocities
profiles for the thin crust end-member, panel b and e for the density dichotomy
crust case, and panels c and f for the thick end-member case.}%
\label{fig:SeisVelsLocal}%
\end{figure}

Depending on the location of seismic events, the propagation of their seismic
waves will encounter not only a different crustal thickness on the path to
the seismic station, but also a different lithospheric thickness and
lithospheric temperature. In Fig.~\ref{fig:SeisVelsLocal} we show
the differences between seismic velocities profiles at present-day at three
different locations on Mars (Tharsis, Utopia, and InSight) for the three
thermal evolution models presented in Fig.~\ref{fig:Cr_HF_Te} and
compare them with their corresponding average profiles. The differences in
the uppermost 400 km can be substantial. The largest differences are observed
between Tharsis volcanic province (profile taken at $-$115$^{\circ}$
longitude and 0$^{\circ}$ latitude) and Utopia impact basin (115$^{\circ}$
longitude and 45$^{\circ}$ latitude), and are most extreme for models with a
thick crust in the Tharsis area (the thick crust model and the model with a
different density between the northern and southern hemispheres).
Interestingly, the model with a different density between the northern and
southern hemispheres shows the largest variations. This is caused by the
higher volumetric heat production in the northern hemisphere compared to the
southern hemisphere in this model. On the other hand, the average profile and 
the profile at the InSight landing site are nearly identical. The largest difference 
between InSight and average profiles is observed for the thickest
crust scenario, where crustal thickness variations between these two locations 
are more pronounced than in the other two crustal thickness models.

The uppermost layers are dominated by the seismic velocities of the crust.
While for the individual profiles at the three locations a sharp transition
occurs at the crust mantle interface, this transition is more gradual for
the average profile. This is due to the fact that for the average profiles
both mantle and crustal areas are present at the same depth. In the uppermost
layers the crustal areas dominate but with deeper depth they become smaller
being replaced by mantle areas. Right below the crust, the seismic velocities
reflect the large variations of the lithospheric temperatures, while deeper
in the convecting mantle these variations become much smaller and therefore
the difference in seismic velocities is minor. The olivine to wadsleyite
phase transition is clearly visible for all models. The depth of the phase
transition depends on the temperature, and for the Tharsis profile, which
has a higher local temperature, the average phase transition depth is at
about ${1110\pm  10}$~km, depending on the thermal evolution model. For
the InSight profile the average phase transition depth is shallower and
about 1070--1080~km, due to a lower temperature at this location. All models
show the same average depth for the olivine to wadsleyite phase transition
at $\sim $1095 km for the average profile, since the latter is nearly identical
for all models (cf. Fig.~\ref{fig:SeisVels}).

In addition to the seismic velocities, the thermal state of the mantle
together with the size and state of the core (solid or liquid, see Section~\ref{sec:Core}) also affects the tidal deformation of a planet. The latter
has been determined for Mars from radio tracking measurements from Mars
Odyssey, Mars Reconnaissance Orbiter, and Mars Global Surveyor
\citep{konopliv16,genova16,konopliv20}. The lag of the tidal deformation
caused by Mars' closest moon, Phobos, is given by the phase lag
$\varepsilon $ that describes the dissipation inside Mars and is linked
to the thermal state of the interior through the viscosity
\citep{nimmo13}. The dissipation can be also expressed in form of the tidal
quality factor $Q$ that is defined as $1/\sin (\varepsilon )$. Low values
of $Q$ would indicate a dissipative mantle caused by a low viscosity and
hence high mantle temperature, whereas a cold interior and consequently
a high viscosity would lead to high $Q$ values.

On Mars, current available estimates for the tidal quality factor
$Q$ calculated at the main tidal period of Phobos (5.55 hours) range between
72 and 105 \citep{ray01,lainey16} and indicate a more dissipative interior
than that of the Earth, for which a tidal quality factor of 280 was calculated
at the lunar semidiurnal terrestrial tide \citep{ray01}. Here we recalculate
the tidal quality factor using the approach from \citet{zharkov05} that
was also used by \citet{khan17}:
%
\begin{equation}
\frac{Q}{k_{2}} = 559
\end{equation}

Using the latest $k_{2}$ estimate of 0.174 $\pm $ 0.008
\citep{konopliv20}, we find a tidal quality factor of $97.3\pm 4.5$. However,
the error bars would increase when accounting for the frequency-dependency
of $k_{2}$, higher tidal terms, and the fact that dissipation occurs in
both Mars and Phobos. Thus, following the approach of \citet{khan17} we
increase the error bars and use a tidal quality factor $Q$ of
$97\pm 12$, a range that includes the previous estimates of
\citet{khan17} as well as the new values obtained from the most recent
$k_{2}$ estimate.

Previous studies have used prescribed thermal profiles
\citep{nimmo13} or temperature profiles from mantle convection models
\citep{plesa18b} to calculate the dissipation in the interior of Mars and
compare the results with observations. The study by \citet{nimmo13} uses
an extended Burgers model for dry olivine and finds that, for a grain size
of 1 cm, a present-day potential temperature of $1625\pm 75$ K is required
to explain the dissipation in the interior of Mars. \citet{plesa18b} used
the present-day thermal state from mantle convection models and computed
the tidal quality factor $Q$. Using $Q$ estimates in the range of
$99.5\pm 4.9$ \citep{konopliv16,lainey16}, this study concluded that models
with an inefficient cooling of the interior caused by either a high amount
of HPEs in the mantle or a large increase of the viscosity with depth (i.e.,
due to a high activation volume) would be too dissipative to satisfy the
constraints. Conversely, models that contain nearly all HPEs in the crust
and have a cold present-day mantle lead to a much lower dissipation than
the suggested values for Mars. Thermal evolution compatible with the
$Q$ estimates of \citet{lainey16} indicates that between 37.6\% and 68.3\%
of the bulk amount of HPEs are concentrated in the crust
\citep{plesa18b}.

\begin{figure}[htbp]%
\centering
\includegraphics[width=0.8\textwidth]{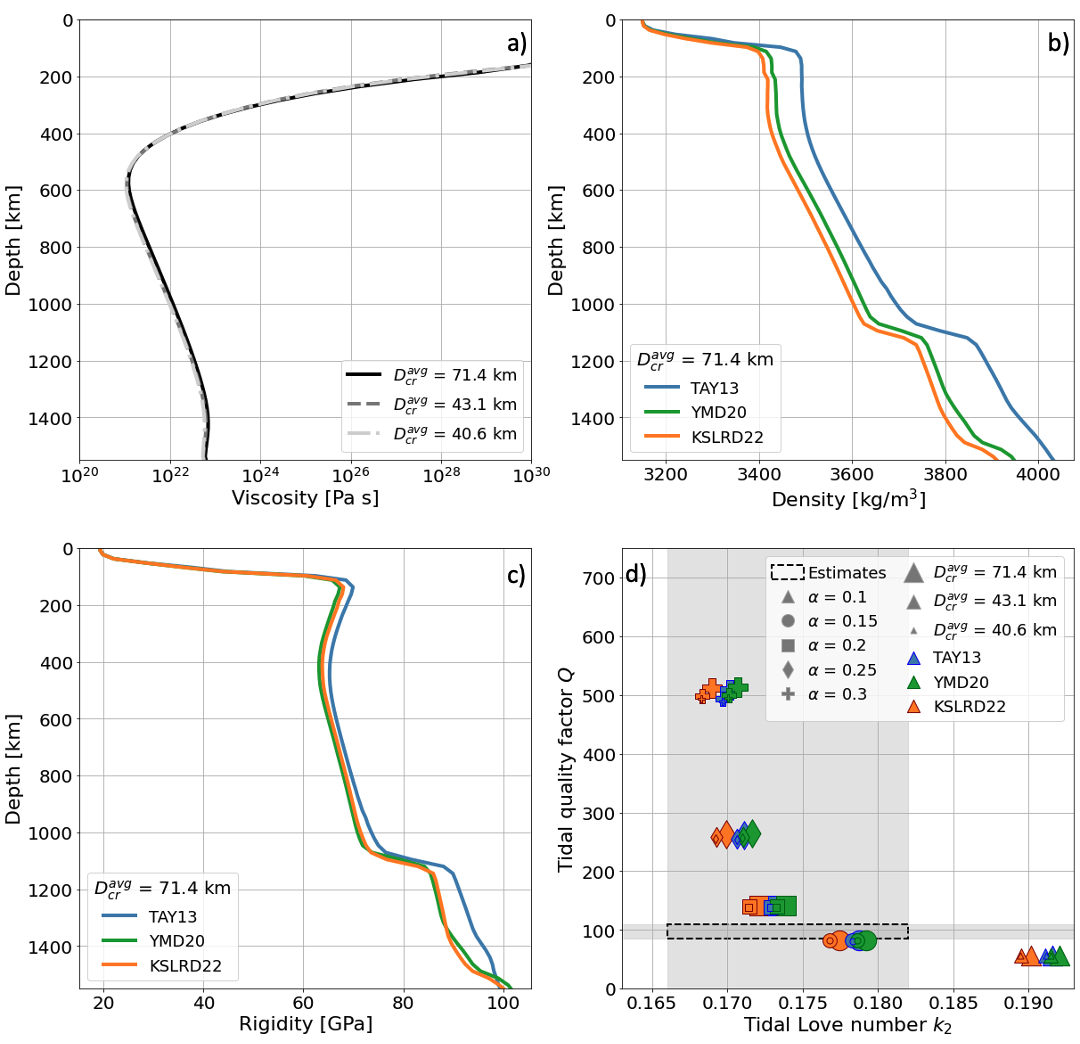}
\caption{Panel a: Viscosity profiles throughout the mantle for the models shown
in Fig.~\ref{fig:Cr_HF_Te}. Panel b and c: mantle density and rigidity profiles,
respectively, for the thick crust end-member using three different mantle
compositions TAY13, YMD20 and KSLRD22. Panel d: calculated tidal quality factor
$Q$ and tidal Love number $k_{2}$ for the three mantle models (panel a) and
three compositions (panel b and c).}%
\label{fig:Q-k2}%
\end{figure}

Here, we calculate the tidal deformation of the three thermal evolution
models presented in Fig.~\ref{fig:Cr_HF_Te}. These models contain 57\%
of the total bulk amount of HPEs in their crust, and thus lie within the
range of models that were found compatible with $Q$ estimates by
\citet{plesa18b}. To compute the tidal deformation, we use a semianalytical
model based on the normal mode theory for radially stratified viscoelastic
bodies \citep{sabadini04}. The results are shown in Fig.~\ref{fig:Q-k2} and Table~\ref{tab:Q-k2}, and are discussed in detail below.

The model uses 100 layers for the mantle and 1 layer for the core, which
is assumes to be homogeneous. It uses as inputs the density, viscosity
and rigidity. While the viscosity profile comes from the geodynamic simulations,
the density and rigidity profiles depend on the mineralogical model and
are calculated by the thermodynamic code Perple\_X \citep{connolly09} using
the thermodynamic formulation and database of \citet{stixrude11}. All three
profiles (i.e., viscosity, density, and rigidity) are temperature and pressure-dependent
and are calculated from the temperature profiles obtained by the geodynamic
models.

\begin{table}[htbp]
\caption{Tidal deformation results for the three models presented in
Fig.~\ref{fig:Cr_HF_Te} using an incompressible Andrade model, three different
mantle mineralogies, and various values of the tidal parameter $\alpha $ used in
the Andrade model. According to \citet{castillo-rogez11} the tidal
parameter $\zeta $ was set to 1 for all calculations.}
\label{tab:Q-k2}
\tabcolsep=4.5pt
\begin{center}
\begin{tabular}{l|cc|cc|cc}
\hline
{\bf Tidal parameter} & \multicolumn{2}{c}{\bf thin crust} & \multicolumn{2}{c}{\bf density dichotomy crust} & \multicolumn{2}{c}{\bf thick crust}\\
\hline
                      & \multicolumn{6}{c}{ TAY13 composition \citep{taylor13}} \\
                      &  $Q$    &   $k_2$  &  $Q$    &   $k_2$  &  $Q$    &   $k_2$\\
\hline
$\alpha = 0.1$        & 56.50   & 0.191 & 56.60   & 0.191  & 56.81   & 0.192 \\
$\alpha = 0.15$       & 81.40   & 0.178 & 81.71   & 0.178  & 82.47   & 0.179 \\
$\alpha = 0.2$        & 137.94  & 0.173 & 138.75  & 0.173  & 140.92  & 0.173 \\
$\alpha = 0.25$       & 253.90  & 0.171 & 255.91  & 0.171  & 261.50  & 0.171 \\
$\alpha = 0.3$        & 489.38  & 0.170 & 494.19  & 0.170  & 508.01  & 0.170 \\
$\alpha = 0.4$        & 1955.25 & 0.169 & 1981.69 & 0.169  & 2060.93 & 0.170 \\
\hline
                      & \multicolumn{6}{c}{ YOS20 composition \citep{yoshizaki20}} \\
                      &  $Q$    &   $k_2$  &  $Q$    &   $k_2$  &  $Q$    &   $k_2$\\
\hline
$\alpha = 0.1$        & 56.76   & 0.191 & 56.86   & 0.191  & 57.09   & 0.192 \\
$\alpha = 0.15$       & 81.90   & 0.179 & 82.21   & 0.179  & 83.00   & 0.179 \\
$\alpha = 0.2$        & 138.98  & 0.173 & 139.81  & 0.173  & 142.03  & 0.174 \\
$\alpha = 0.25$       & 256.20  & 0.171 & 258.24  & 0.171  & 263.93  & 0.172 \\
$\alpha = 0.3$        & 494.50  & 0.170 & 499.37  & 0.170  & 513.42  & 0.171 \\
$\alpha = 0.4$        & 1980.81 & 0.169 & 2007.58 & 0.170  & 2088.06 & 0.170 \\
\hline
                      & \multicolumn{6}{c}{ KHA22 composition \citep{khan22}} \\
                      &  $Q$    &   $k_2$  &  $Q$    &   $k_2$  &  $Q$    &   $k_2$\\
\hline
$\alpha = 0.1$        & 56.69   & 0.189 & 56.79   & 0.189  & 57.01   & 0.190 \\
$\alpha = 0.15$       & 81.77   & 0.177 & 82.08   & 0.177  & 82.87   & 0.177 \\
$\alpha = 0.2$        & 138.72  & 0.171 & 139.53  & 0.171  & 141.74  & 0.172 \\
$\alpha = 0.25$       & 255.58  & 0.169 & 257.60  & 0.169  & 263.27  & 0.170 \\
$\alpha = 0.3$        & 493.03  & 0.168 & 497.87  & 0.168  & 511.85  & 0.169 \\
$\alpha = 0.4$        & 1972.59 & 0.168 & 1999.18 & 0.168  & 2079.16 & 0.168 \\
\hline
\end{tabular}
\end{center}
\end{table}

For the tidal deformation calculations, we used the Andrade rheological
model and assumed that the planet is incompressible. Nevertheless, we note
that the assumption of compressibility might render corrections to
$k_{2}$ that are of the order of the observational uncertainty. While other
rheological models exist and have been applied to calculate the tidal deformation
of Mars and other rocky planets, the advantage of the Andrade model is
the small number of parameters it requires for the calculations
\citep{castillo-rogez11,efroimsky12}. A detailed review of the theory
of viscoelasticity and tidal response, as well as their application to
constrain the interior structure of Mercury, Venus, Mars, the Moon, and
icy satellites is given in \citep{bagheri22}.

A recent study by \citet{bagheri19} has compared various rheological models
for Mars and found that all models can fit the observations when using
a single frequency (i.e., at the main period of Phobos), but information
of the dissipation at additional frequencies could help to distinguish
between the current rheological models. The same study concluded that the
Maxwell rheology would require very low viscosities to fit the available
data and rheologies such as Andrade, extended-Burgers, or Sundberg-Cooper
are more appropriate to use when studying tidal dissipation.

The Andrade model that is used here is able to describe all components
of deformation (elastic deformation, viscous creep, and the transient Andrade
creep) and requires in total four parameters: the viscosity, the rigidity
and two empirically determined parameters $\alpha $ and $\zeta $. The parameter
$\zeta $ describes the ratio between the timescales of the anelastic Andrade
creep and the Maxwell body and has been found to be close to one
\citep{castillo-rogez11}. On the other hand $\alpha $ describes the duration
of the transient response, and values for olivine-rich mantle rocks lie
between 0.1 and 0.5, and mostly between 0.2 and 0.4
\citep{castillo-rogez11}.

The average thermal state of the geodynamic models, which was used to calculate
the tidal deformation in Fig.~\ref{fig:Q-k2}, is very similar (cf. Fig.~\ref{fig:SeisVels}a). Although the temperature and seismic velocities variations
in the lithosphere can be significantly different between the three geodynamic
models (cf. Fig.~\ref{fig:SeisVelsMaps}), the average viscosity, density,
and rigidity profiles are very similar (Fig.~\ref{fig:Q-k2}a, b, and c),
with TAY13 composition showing slightly larger densities and higher rigidities
than YOS20 and KHA21 composition, due to the higher FeO content. Thus,
the tidal dissipation values mainly depend on the chosen value for
$\alpha $, for which a value between 0.2 and 0.15 seems to fit best the
observed dissipation in the interior of Mars (Fig.~\ref{fig:Q-k2}d).

\section{Core radius estimates and their implications for the interior dynamics}
\label{sec:Core}

The core of a terrestrial planet is a witness of the earliest planetary
differentiation, when metal and silicates separate to form the layers inside
the planet. The size of the core is essential to determine the thickness
of the silicate layer (mantle and crust), when knowing the planet's radius.
The thickness of the silicate layer in turn affects the mantle flow and
the convection pattern (i.e., number of mantle plumes and their distribution).
The latter can be linked to surface geological features such as volcanic
and tectonic provinces. For a detailed review that describes the methods
and the progress in determining of the core size of the Earth, Mars, and
Moon, as well as future opportunities for terrestrial exoplanets we refer
the reader to \citep{knapmeyer22}.

Many geodynamic studies have investigated the formation of the martian
crustal thickness dichotomy, proposing a degree-one or ridge like convection
pattern. This pattern is largely favored for models using a small core
that allows for the presence of an endothermic phase transition at the
base of the mantle \citep{harder96,breuer98}, similar to the 660-phase
transition on the Earth. Models employing a specific mantle viscosity structure
with a viscosity increase in the mid-mantle
\citep{zhong01,roberts06,keller09} and models that included the combined
effects of a giant impact and the subsequent dynamics in the mantle
\citep{golabek11} were also able to produce a degree-one mantle pattern,
but in these cases too the core radius was about half of the planetary
radius or smaller.

InSight's measurements have revealed that the martian core has a radius
of $1830\pm 40$~km and is more than half the planet's radius
\citep{staehler21}. This is consistent with estimates of the tidal Love
number $k_{2}$ \citep{konopliv16,genova16} that were previously combined
with thermal evolution models and suggested a core radius strictly larger
than 1800 km \citep{plesa18b}. The $k_{2}$ value of
\citet{konopliv16} of $0.169\pm 0.006$ has been recently updated by
\citet{konopliv20} to $0.174\pm 0.009$. The most recent estimate was corrected
for atmospheric tides, and the uncertainties account for the fact that
the correction for atmospheric tides depends on the atmospheric conditions
at the time of observations \citep{konopliv20}. As shown in Fig.~\ref{fig:Q-k2} models using a core radius of 1850 km, a value that was
previously used by \citet{plesa18b} and is consistent with the seismic
detection of the martian core \citep{staehler21}, are also able to fit
the latest $k_{2}$ estimate of \citep{konopliv20}.

\begin{figure}[htbp]%
\centering
\includegraphics[width=\textwidth]{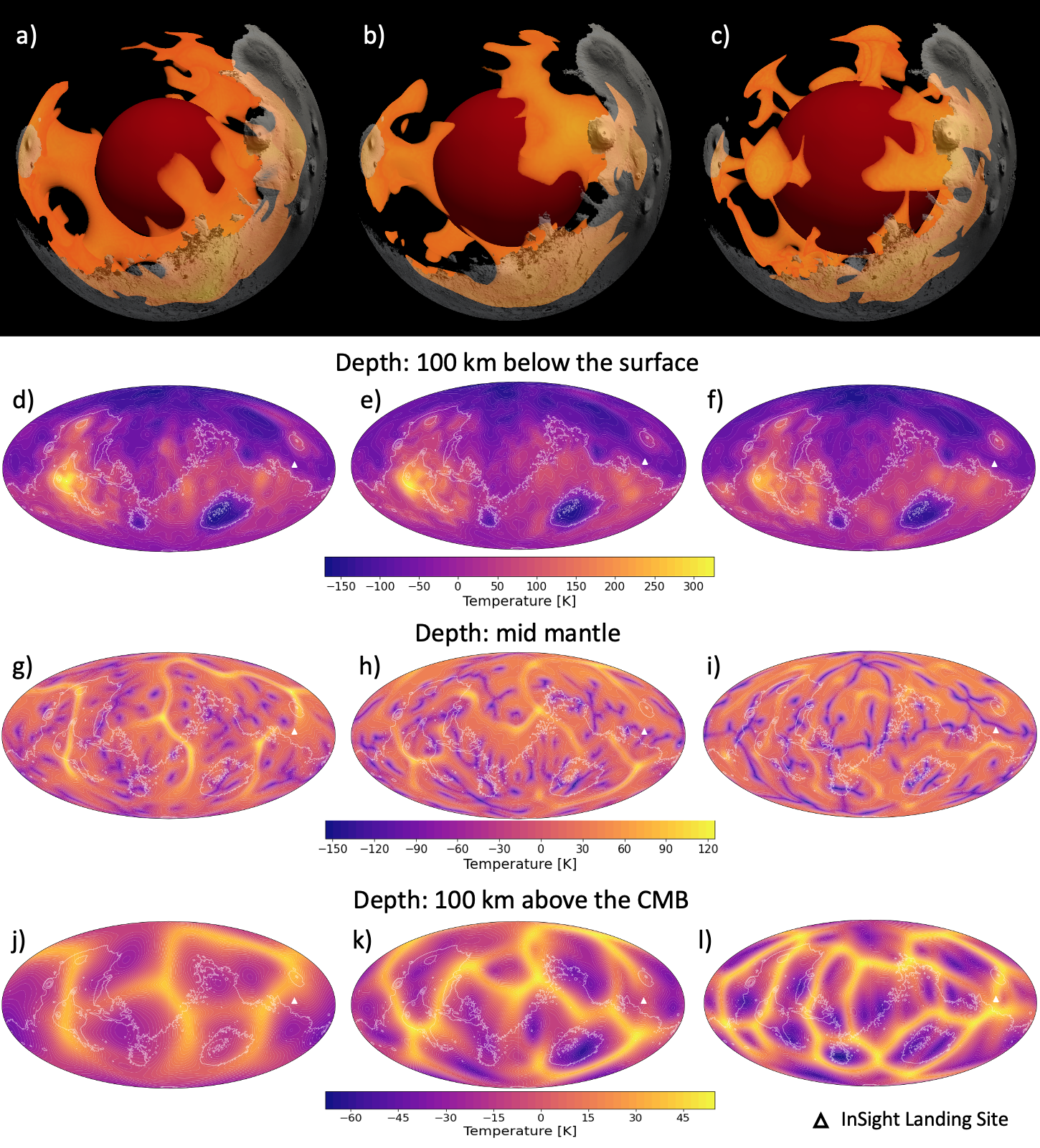}
\caption{Mantle convection pattern (panels a, b, and c) and temperature
variations throughout the mantle. All models use a crustal thickness with and
average value of 61.3 km and a crustal density of 2800 kg\,m$^{-3}$. Left column
(panels a, d, g, and j) shows a model with a core radius of 1500 km. Middle
column (panels b, e, h, and k) shows a model with a 1700 km core radius, while
right column (panels c, f, i, and l) presents a model with a core radius of 1850
km. Only the model shown in the right column is compatible with core radius
estimates by InSight.}%
\label{fig:Pattern}%
\end{figure}

The large size of the core excludes the possibility of having a bridgmanite-dominated
lower mantle on Mars \citep{staehler21}, as it is the case for the Earth.
An endothermic phase transition at the base of the martian mantle will
no longer occur, as the pressure is too low for this to take place. Even
for models with smaller core radii (1700 -- 1360 km) an endothermic phase
transition at the base of the present-day martian mantle was only marginally
possible requiring CMB temperatures in excess of 2100 K
\citep{spohn98}. In addition, the large radius of the core leads to a small
scale convection pattern with many small plumes distributed throughout
the mantle as illustrated in Fig.~\ref{fig:Pattern}.

All three models in Fig.~\ref{fig:Pattern} have been built similarly to
the models presented in \citep{plesa16}. They use the same crustal thickness
with a crustal density of 2800 kg\,m$^{-3}$ and a crustal enrichment that
matches the average value derived from GRS measurements. The crustal thickness
is derived from gravity and topography data and matches the crust-mantle
discontinuity at InSight landing site, as observed in the seismic measurements
\citep{knapmeyer-endrun21,wieczorek22}. The radius of the core has been
varied between the three models with values of 1500 km, 1700 km, and 1850
km. While the former two values are incompatible with the recent InSight
data, the last value of 1850 km lies well within the current core radius
estimates \citep{staehler21}.

The models show a shallow subsurface that is mostly dominated by the crustal
thickness variations with a temperature pattern similar to the crustal
thickness pattern (Fig.~\ref{fig:Pattern}d--f). The effects of
mantle plumes are more pronounced for 1500 km core radius compared to the 1850
km core radius. It can be observed that the case with the largest core shows
a short wave-length convection pattern compared to the small core model. This
is illustrated both in the temperature maps at mid-mantle depth
(Fig.~\ref{fig:Pattern}g--i) and at 100 km above the CMB
(Fig.~\ref{fig:Pattern}j--l) that show the presence of a larger
number of plumes and downwellings for the model with a core radius of 1850 km
compared to the case with a core radius of only 1500~km.\looseness=1

A large core that leads to a small scale mantle convection pattern is at
odds with the formation of the martian crustal thickness dichotomy through
an endogenous process. However, exogenic processes such as the sequence
of one or several large impacts may represent a key mechanism
\citep{wilhelms84,frey88}. Such scenario is compelling, as it has been
proposed to explain the elliptical nature of the dichotomy
\citep{andrews-hanna08,nimmo08,marinova08}. The effect on the subsequent
dynamics in the mantle may lead to mantle plumes in Tharsis and Elysium
that with time may be stabilized by the insulating effect of a thicker
crust at those locations~\citep{schumacher06}. This scenario, however, would
not exclude weaker thermal anomalies at other locations, that may have
led to shorter episodes of volcanic activity (e.g., Syrtis Major, the Circum
Hellas province) or may have never resulted in the buildup of volcanic
provinces at the surface.

Another important aspect of a large martian core is that it requires a
high amount of light elements to be able to match the planet's mass. Sulfur
cannot be the only light element in the core, as its amount would then
exceed the abundance found in EH-chondrites, the most sulfur rich building
blocks. Thus, other elements such as oxygen, carbon, and hydrogen are required
to match both the core density and to be compatible with geochemical arguments
\citep[][and references therein]{staehler21,khan22}. The composition of
the martian core has major consequences for its evolution and the generation
of an early magnetic field. A large amount of light elements in the core
could lead to a scenario in which the core crystallizes from the top down
(iron snow) by forming iron particles at the top of the core due to a steeper
melting temperature than the core adiabatic profile
\citep{stewart07,rivoldini11,breuer15iron,helffrich17,davies18,hemingway21}.

A recent study by \citet{hemingway21} investigated the crystallization
of the martian core and found that Mars may possess a partially solid core
today. However, the large amount of light elements places the core composition
close to the eutectic and likely prevented the crystallization of an inner
core \citep{staehler21}, due to the significant decrease of the core melting
temperature \citep{mori17}. In addition, \citet{hemingway21} used a simple
thermal evolution model without explicit treatment of the stagnant lid
evolution. This most likely underestimates the mantle and core temperatures.
Typical stagnant lid thermal evolution models
\citep[e.g.,][]{breuer03,morschhauser11,plesa18b,samuel19,samuel21}
suggest that even at present day the CMB temperature is too high to allow
for core crystallization, in particular for a high mantle viscosity that
was found compatible with additional geophysical and seismic constraints
\citep{plesa18b,samuel19,knapmeyer-endrun21}. However, core crystallization
would undoubtedly take place in the future, when the core temperature has
sufficiently decreased to allow for core crystallization. The exact time
and style of crystallization will strongly depend on the core composition
and thermal state \citep{stewart07}.

The early martian dynamo was most likely driven by thermal buoyancy inside
the core until at least 3.7 Gyr ago \citep{mittelholz20}, placing important
constraints on the heat flow at the CMB. Thermal evolution scenarios that
maintain a CMB heat flow above the critical core heat flow, above which
thermal convection in the core sets in, for at least 800 Myr need to be
investigated in future studies. An important step for answering this question
has been undertaken by \citet{greenwood21}, who used 1D parametrized thermal
evolution models and showed that a prolonged thermally driven dynamo can
be sustained. Successful models require core thermal conductivities in
the range of 16 -- 35 W\,m$^{-1}$\,K$^{-1}$ and mantle reference viscosities
of 10$^{21}$ Pa\,s or smaller and activation volumes smaller than 6 cm$^{3}$\,mol$^{-1}$
\citep{greenwood21}. However, geodynamic models in a 3D geometry require
higher viscosities and/or activation volumes to explain localized melting
at recent times \citep{knapmeyer-endrun21} and a large present-day elastic
lithosphere thickness at the north pole of Mars \citep{plesa18b}. Thus
future models need to investigate if both constraints for the early magnetic
field and recent thermal state can be matched.

\section{Seismogenic layer thickness and the present-day seismicity}
\label{sec:Seismicity}

Seismic observations provide the most direct view into the interior of
a planetary body and reveal the level of activity that the planet experiences.
While seismic observations have greatly improved our understanding of the
interiors of the Earth, Moon, and Mars, currently, the seismic activity
for other planetary bodies such as Mercury, Venus, or icy satellites can
only be indirectly estimated with large uncertainties. A comprehensive
review of the current state of knowledge of planetary seismology and directions
for future seismic investigations of planetary bodies is given in \citep{staehler22}.

In the absence of plate tectonics and because Mars is smaller than the
Earth, its seismicity was suggested to be lower than that of the Earth
and mainly driven by planetary cooling. However, the level of seismic activity
was expected to be larger than the seismicity of the Moon recorded by the
Apollo seismic measurements \citep{ewing71,toksoz74}. Previous seismic
measurements on Mars were performed by Viking in the 1970s
\citep{anderson76}. While the seismometer on Viking 1 failed to uncage
and could not record any data, Viking 2 collected data between 1976 and
1978. However, these measurements are strongly contaminated by noise caused
by lander vibrations due to wind, given the location of the seismometer
on the lander deck. During 146 sols of operation only one event recorded
by Viking 2 seismometer could be interpreted as a marsquake
\citep{anderson77}, but a seismic origin is difficult to establish
in the absence of wind data during the event.

In the absence of unambiguous seismic recordings from Mars, previous studies
have estimated the level of seismic activity based on the analysis of surface
faults \citep{golombek92,golombek94,golombek02} and from numerical models
of planetary cooling \citep{phillips91,knapmeyer06,plesa18}. Maps of tectonic
centers were compiled based on orbital imaging of the surface. Using the
Mars Orbiting Laser Altimeter shaded topographic relief maps,
\citet{knapmeyer06} compiled a global fault catalog and used it to predict
the martian seismicity and the distribution of epicenters by associating
the event size with fault length. Another study by \citet{plesa18} used
3D geodynamic thermal evolution models combined with spatial variations
of crustal thickness to evaluate the seismogenic layer thickness and the
present-day martian seismicity. While all previous models predicted that
Mars is seismically active today with a seismicity between that of the
Moon and that of the Earth, the uncertainties of the annual seismic moment
covered several orders of magnitude.

Since more than three years, InSight's seismometer (SEIS) has been recording
seismic events on Mars. In the absence of microseismic events that are
observed on the Earth, SEIS is able to record extremely small amplitude
events on Mars \citep{lognonne20}. Although sensitive to the martian wind
that leads to a noisy environment during the martian mid-day
\citep{giardini20}, SEIS was able to record over 2000 teleseismic events
\citep{mqs22} mostly during the late afternoon and evening, when the noise
level is low. For some of these events the location could be determined
and several of them have been localized in the Cerbeus Fossae region
\citep{zenhausern22} -- a young fault system with a minimum age of 10 Myr
situated between 20$^{\circ}$ and 40$^{\circ}$ east of the InSight landing
site \citep{taylor13Cerberus}. A recent study by \citep{horleston22} reported
on distant seismic events, one of which could be located in Valles Marineris
(146$^{\circ}\pm $7$^{\circ}$). For many events, however, a localization
is difficult in particular due to large uncertainties in the backazimuth
that are caused by high scattering and noise levels in the seismic data.
Therefore, for many of the recorded seismic events only a distance can
be provided, while the direction that is required to determine the location
of the source remains unclear \citep{giardini20}. Recent advances in the
study of these events, by using a comprehensive polarization analysis,
have been applied to improve the estimates of the distribution of seismicity
on Mars \citep{zenhausern22}.

\begin{figure}[htbp]%
\centering
\includegraphics[width=\textwidth]{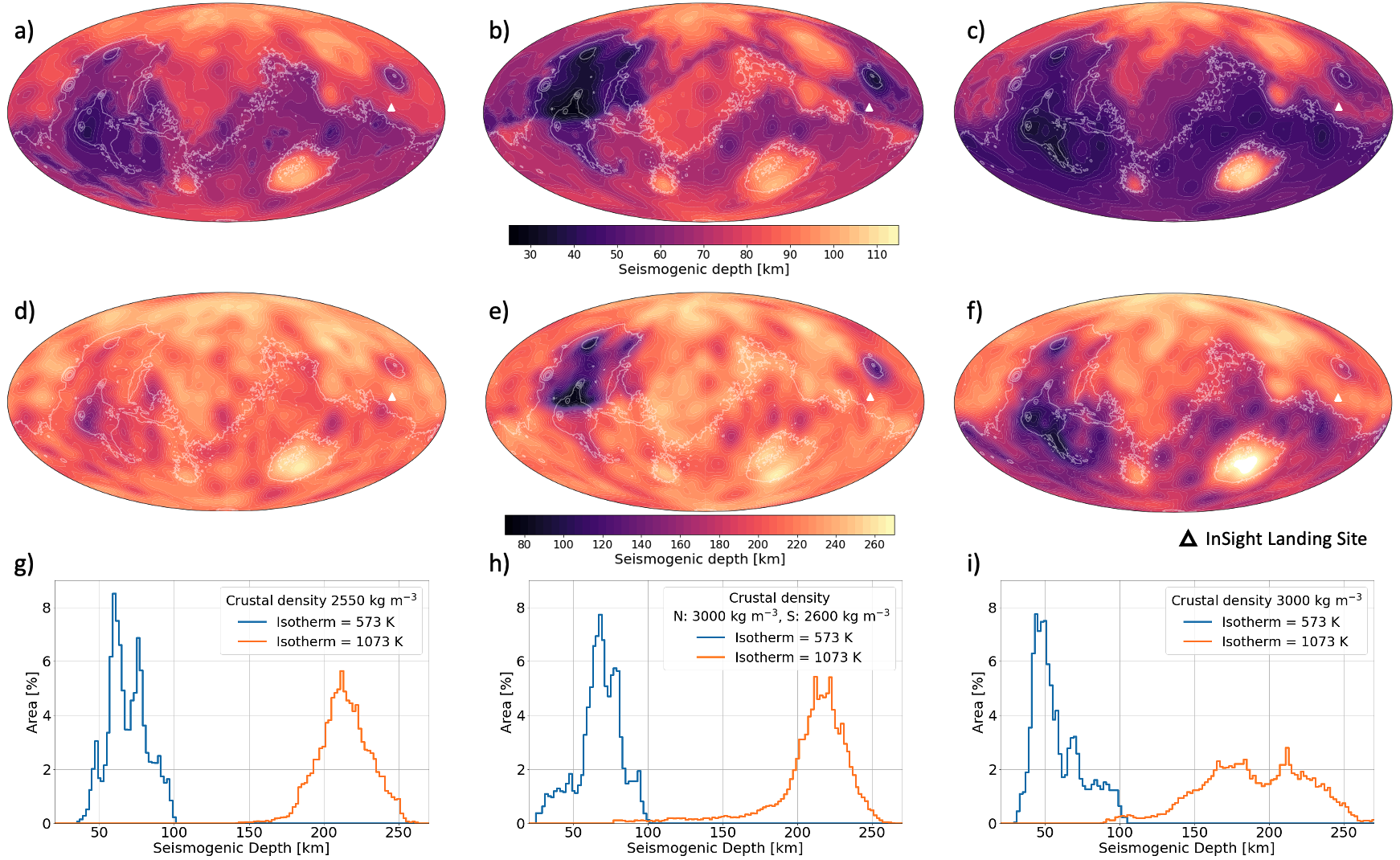}
\caption{Maps of the seismogenic layer at present day computed using the 573 K
isotherm (panels a, b, and c) and the 1073 K isotherm (panels d, e, and f).
Panels g, h, and i show histograms for the seismogenic layer thickness. Panels
a, d, and g show the results obtained for the thin crust end-member case, Panels
b, e, and h show the seismogenic layer thickness for the density dichotomy
crust, and panels c, f, and i show the values that were obtained for the thick
crust end-member. For orientation, the white contour lines show the 0 km level
of surface topography.}%
\label{fig:SeisDepth}%
\end{figure}

On Mars planetary cooling was thought to be the main source of present
day seismicity. However, high frequency events detected by InSight may
be driven by solar illumination, the CO$_{2}$ cycle or annual solar tides
\citep{knapmeyer21}. Moreover, the high level of seismicity observed in
Cerberus Fossae \citep{zenhausern22} could be indicative of processes such
as magma ascent thorough the crust and lithosphere that may be ongoing
on Mars. Indeed, some of the low frequency marsquakes have been suggested
to be related to volcanic tremor in Elysium Planitia region
\citep{kedar21}. Thus, seismicity may not only be linked to the cooling
of the interior, but also to ongoing magmatic processes, and knowledge
about the level of seismic activity and location of seismic events can
help to constrain the evolution and present-day state of the martian mantle.
Global thermal evolution models which use crustal thickness variations
derived from gravity and topography data and anchored by the seismic observations
at InSight landing site, show a close correlation between the crustal thickness
variations and the seismogenic layer thickness (Fig.~\ref{fig:SeisDepth}). The latter is typically estimated by using an isotherm
that describes the depth up to which seismic events could originate. In
previous studies isotherms between 573 K and 1073 K have been tested
\citep{phillips91,knapmeyer06,plesa18}. The 573 K isotherm marks the temperature
at which quartz, the most ductile component of a granitic crust starts
to show a plastic behavior \citep{scholz98}. Thus, this temperature is
often associated with the bottom of the seismogenic layer on Earth. The
1073 K isotherm is more representative for a basaltic composition, as it
marks the maximum depth of oceanic intraplate quakes
\citep{bergman86,wiens83}. Since the majority of the martian crust is thought
to have a basaltic composition, this value has been suggested to be more
representative for determining the depth of the seismogenic layer on Mars.

In Fig.~\ref{fig:SeisDepth} the depth of the seismogenic layer is
shown for the three thermal evolution models discussed in
Section~\ref{sec:Lith} that employ the crustal thickness variations
illustrated in Fig.~\ref{fig:Cr_HF_Te}. The seismogenic layer was
calculated by using the 573 K (Fig.~\ref{fig:SeisDepth}a,
b, and c) and 1073 K isotherms (Fig.~\ref{fig:SeisDepth}d, e, and
f), and the results are shown in Table~\ref{tab:SeisLayer}. Similar
to the models presented in \citet{plesa18}, due to their effects on the
lithospheric temperatures, the variations of the crustal thickness control
the seismogenic layer thickness variations. For the 573 K isotherm a clear
dichotomy can be observed for the seismogenic layer thickness of the thin and
thick crust end-members (Fig.~\ref{fig:Cr_HF_Te}a and b). For the
density dichotomy crust, where variations in crustal density reduce the
variations in crustal thickness, the seismogenic layer thickness is more
homogeneous, but shows small values in Tharsis and Elysium provinces that are
characterized by a thicker crust compared to the rest of the planet. For the
1073 K isotherm the crustal thickness dichotomy pattern is no longer visible
for the thin crust model, as in this case this temperature is attained at a
depth that is no longer sensitive to the temperature variations caused by the
crustal thickness pattern. Due to the thicker crust and higher amount of
crustal HPEs in the thick crust model, a dichotomy in the seismogenic layer is
still visible for the 1073 K isotherm. The crustal density dichotomy model
shows a thin seismogenic layer in Tharsis and Elysium areas and otherwise a
rather homogeneous distribution.

\begin{table}[htbp]
\caption{Seismogenic layer thickness obtained the three models presented in
Fig.~\ref{fig:Cr_HF_Te} using a 573 K and a 1073 K isotherm. Min, median, and
max show the minimum, median and maximum values attained in each model.}
\label{tab:SeisLayer}
\begin{center}
\begin{tabular}{l|c|ccc}
\hline
{\bf Isotherm}          &        & {\bf thin crust} & {\bf density dichotomy crust} &  {\bf thick crust}\\
                        &        &      [km]        &           [km]                &      [km]\\
\hline
\multirow{3}{*}{573 K}  & min    & 38.77            &    28.71                      &    33.74 \\  
                        & median & 76.40            &    75.83                      &    58.52 \\
                        & max    & 104.16           &    104.16                     &    109.19\\
\hline
\multirow{3}{*}{1073 K} & min    & 143.42           &    79.48                      &    90.07 \\
                        & median & 219.15           &    221.50                     &    193.87\\
                        & max    & 265.29           &    266.55                     &    281.77\\
\hline
\end{tabular}
\end{center}
\end{table}

For the 573 K isotherm the seismogenic depth is much shallower compared
to the 1073 K isotherm. The seismogenic layer thickness of the models presented
in Fig.~\ref{fig:SeisDepth} show values between 29 and 109 km when the
seismogenic layer is defined using the 573 K isotherm, while for the 1073
K isotherm the seismogenic layer thickness extends to depths of 282 km.
Compared to the 573 K isotherm, the range of seismogenic layer
thicknesses obtained with the 1073 K isotherm is nearly twice as large.
As already shown by \citet{plesa18} the range of seismogenic layer thickness
increases for large crustal thickness variations. The values presented
in Fig.~\ref{fig:SeisDepth} are smaller than the high density crust models
(HC models) of \citet{plesa18}, but similar to the values obtained for
the crustal thickness of \citet{neumann04} (NC models) and those for a
density dichotomy crust (DC models). This is due to the fact that the HC
models of \citet{plesa18} have a thick crust (87.1 km on average) with
a high amount of crustal HPEs and a cold mantle and lithosphere that lead to 
a thick seismogenic layer. Later, these models were excluded, as they were found to produce
a much thinner elastic thickness than the south pole estimate due to the
presence of a decoupling layer between the elastic cores of the mantle
and crust \citep{plesa18b}. Furthermore, the average thickness of the HC models
exceeds 72 km and is thus incompatible with the recent seismic data of the InSight
mission \citep{knapmeyer-endrun21}.

The seismogenic layer thickness is directly linked to the depth of seismic
events, since deep seismic events could be indicative of a cold and thick
lithosphere. For the three models shown in Fig.~\ref{fig:SeisDepth} some
regions such as the Tharsis region, Elysium Planitia, and Arabia Terra
region show a different seismogenic thickness depending on the exact model
and the isotherm used. We note, however, that deep events are necessary
to be able to distinguish between seismogenic depth distributions predicted
from thermal evolution models \citep{plesa18}. Shallow seismic events can
occur in both thin and thick seismogenic layers. In general, deep seismic
events would indicate that the 1073 K isotherm is more appropriate to define
the seismogenic layer thickness. Specifically, deep seismic events in e.g.,
Arabia Terra could be indicative of a seismogenic thickness distribution
such as the one observed for the density dichotomy crust (Fig.~\ref{fig:SeisDepth}), while deep seismic events in Daedalia Planum around
Arsia Mons and in the adjacent Terra Sirenum region could exclude the thick
crust model. Deep seismic events in the northern part of the Tharsis province
around Alba Mons would favor the thin crust model, while seismic events
as deep as 280 km could only be obtained in Hellas basin in the thick crust
model. However, source depths of marsquakes recorded by InSight lie at
about 20--50 km below the surface \citep{brinkman21,staehler21}, but depth
uncertainties remain large \citep{brinkman21}. Thus, currently none of
the seismogenic layer distributions suggested by \citet{plesa18} and shown
in Fig.~\ref{fig:SeisDepth} can be excluded based on the depth of seismic
events.

The seismogenic layer volume can be calculated from the seismogenic layer
thickness and can be used to estimate an annual seismic moment knowing
the strain rate from thermal evolution models.
%
\begin{equation}
\label{eq_mcum}
M_{cum} = \eta \dot{\varepsilon} V \mu \Delta t,
\end{equation}
where $\eta $ is the seismic efficiency with values between 0 and 1 that
describe how much of the strain is released in form of seismic events compared
to aseismic deformation. The strain rate $\dot{\varepsilon}$ is estimated
from thermal evolution models. $V$ is the seismogenic layer volume,
$\mu $ is the shear modulus and $\Delta t$ is the time interval used to
compute the seismic moment.

Previous studies by \citet{knapmeyer06} and \citet{phillips91} have used
parametrized thermal evolution models and investigated the rate of planetary
cooling to estimate an annual seismic moment. In a more recent study,
\citet{plesa18} used global 3D models and estimated the annual seismic
moment distribution based on the local contributions of strain rates associated
with mantle cooling and convection. The contribution associated with convective
stresses was found to be high in regions covered by a thick crust that
leads to higher subsurface mantle temperatures and lower viscosities allowing
for material to flow. The contribution associated with cooling stresses
on the other hand was found to be high in area covered by a thin crust
such as the northern hemisphere or large impact basins. In the absence
of a thick insulating crust, these areas cool more efficiently and can
produce higher cooling stresses compared to regions covered by a thick
crust. While the seismic moment contributions from convective and cooling
stresses are anti-correlated, given the fact that Mars is a stagnant lid
planet and thus convective stresses are negligible in the shallow subsurface,
the contribution from convective stresses is typically smaller compared
to that of cooling stresses. Moreover, the contribution of convective stresses
is entirely absent for the 573 K isotherm, as this isotherm would lead
to a thin seismogenic layer.

\begin{figure}[htbp]%
\centering
\includegraphics[width=\textwidth]{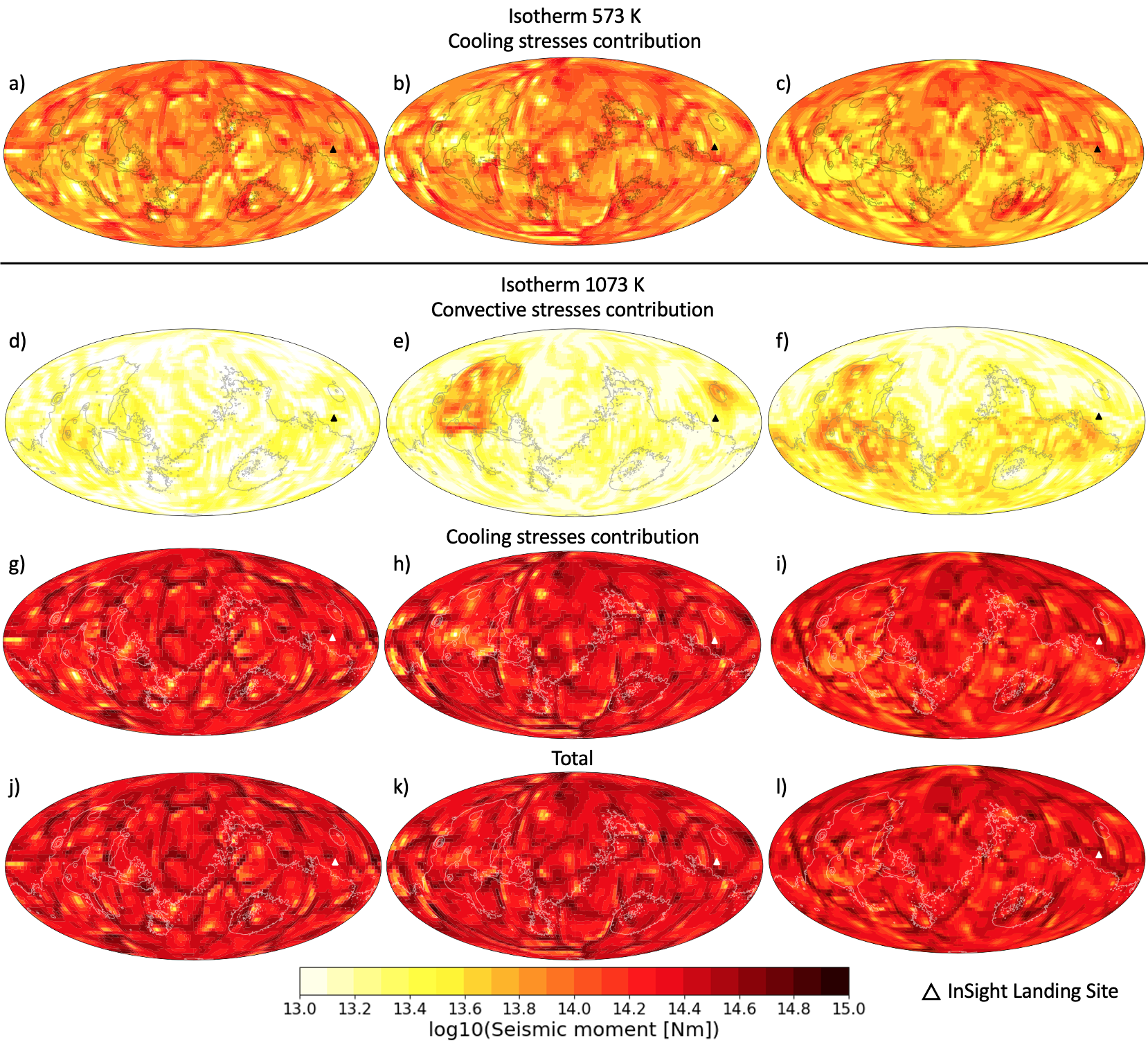}
\caption{Spatial distribution of the annual seismic moment budget computed using
the 573 K isotherm to define the seismogenic layer thickness (panels a, b, and
c). Panels d, e, and f show the convective stresses contribution, panels g, h,
and i the contribution associated with cooling stresses, and panels j, k, and l
the total annual seismic moment budget computed using the 1073 K isotherm. Left
column (panels a, d, g, and j) shows the results obtained for the thin crust
end-member case, middle column (panels b, e, h, and k) shows the case with a
density dichotomy crust, and right column (panels c, f, i, and l) shows the
thick crust end-member case. The white and gray contour lines show the 0 km
level of surface topography.}%
\label{fig:SeisMom}%
\end{figure}

The distribution of the annual seismic moment is shown in Fig.~\ref{fig:SeisMom}. For the 573 K isotherm the distribution of the annual
seismic moment reflects the cooling pattern of the lithosphere that is
controlled by the crustal thickness variations. Areas covered by a thick
crust show a lower annual seismic moment budget, due to their slower cooling
compared to areas covered by a thin crust. For the 1073 K isotherm, on
the other hand, the annual seismic moment distribution is rather homogeneous.
The contribution associated with convective stresses illustrates that a
higher seismic moment can be attained in the southern hemisphere and the
Tharsis and Elysium volcanic provinces, given their thicker crust that
leads to warmer temperatures in those regions. The cooling stresses are
more homogeneously distributed with slightly lower values in Tharsis and
the southern hemisphere in particular for the density dichotomy and thick
crust models (Fig.~\ref{fig:SeisMom}h and i, respectively). Nevertheless,
when combining the contributions from cooling and convective stresses,
the total contribution leads to a more homogeneous pattern, as it was also
discussed by \citet{plesa18}.

\begin{figure}[htbp]%
\centering
\includegraphics[width=\textwidth]{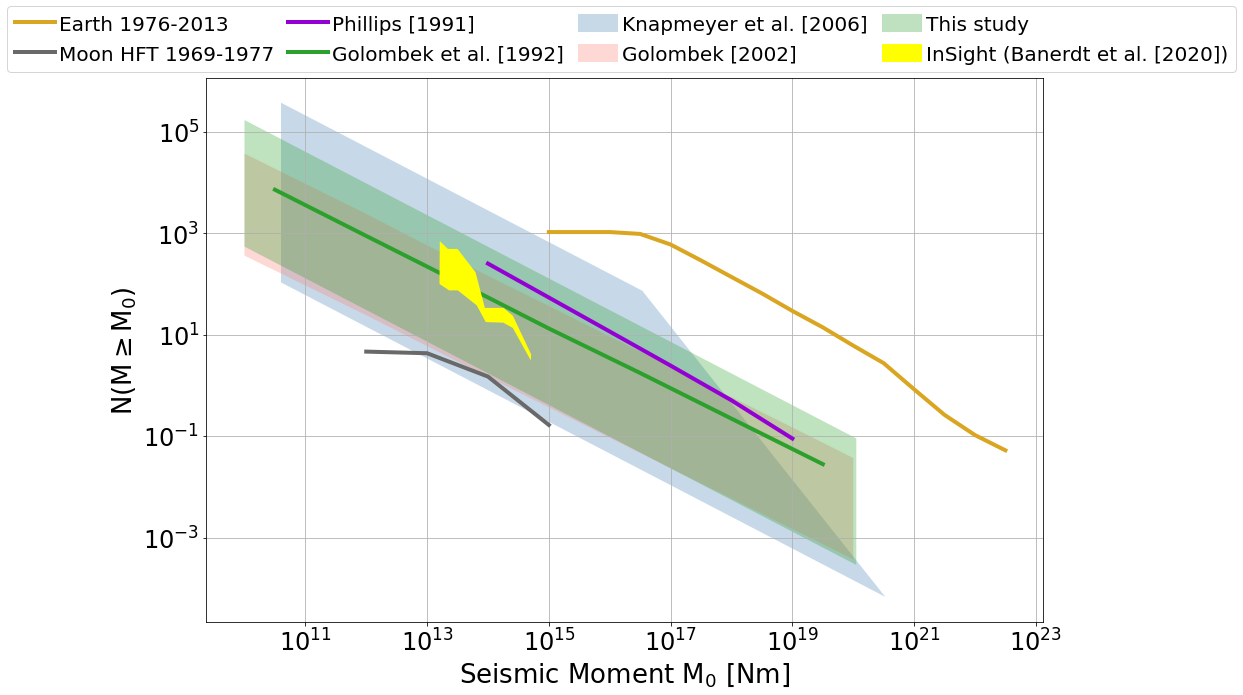}
\caption{Comparison of the moment-frequency diagram for the Earth, Moon, and
Mars. For Mars, the comparison includes the models presented in
Fig.~\ref{fig:Cr_HF_Te} (this study), previous seismicity estimates from
\citet{knapmeyer06}, \citet{golombek02}, \citet{golombek92}, and
\citet{phillips91}, as well as the values derived from the InSight data
\citep{banerdt20}. Similar to \citet{plesa18}, the maximum seismic moment for a
marsquake was assumed to be $10^{20}$ Nm and the slope was set to 0.625
\citep{knapmeyer06}, as suggested from the analysis of quakes occurring above the
olivine-bridgmanite transition on Earth \citep{kagan02}.}%
\label{fig:Seismicity}%
\end{figure}

The total available annual seismic moment budget can be used to compute
a size-frequency distribution that often follows a Gutenberg-Richter law.
The size-frequency distribution indicates the number of events that would
be expected to occur over the course of a year with a seismic moment larger
or equal to the largest assumed marsquake (seismic moment M\_0). The moment
release obtained from global 3D thermal evolution models that include the
recent constraints from the InSight data on core size and crustal thickness
is shown in Fig.~\ref{fig:Seismicity}. Models indicate an annual cumulative
seismic moment between $5.19\times 10^{16}$ and $1.52\times 10^{19}$ Nm,
similar to previous values from thermal evolution calculations
\citep{knapmeyer06,plesa18}. Fig.~\ref{fig:Seismicity} also includes previous
estimates from lithospheric cooling computed using parametrized thermal
evolution models \citep{knapmeyer06,phillips91}, as well as from total
slip on surface faults \citep{golombek92,golombek02}. The seismic moment
release of the Earth was obtained from Harvard Centroid Moment Tensor catalog
between 1976 and 2013, while that of the Moon was derived from shallow
moonquakes \citep{oberst87}. The seismicity of Mars was derived based on
InSight observations of marsquakes and extrapolated to the entire planet
to account for uncertainties in detecting small and distant events
\citep{banerdt20}.

The size-frequency diagram in Fig.~\ref{fig:Seismicity} is sensitive to
the largest seismic moment assumed. Values of the maximum possible seismic
moment for Mars were estimated based on data from oceanic and continental
intraplate quakes on Earth and lie around $10^{20}$ Nm\,yr$^{-1}$
\citep{phillips91,golombek92,golombek94}. Adopting this value and using
the total cumulative moment release (Eq.~(\ref{eq_mcum})) the moment-frequency
relation can be calculated. Additional uncertainties are related to the
seismic efficiency $\eta $ that lies between 0.025 and 1 for events on
Earth \citep{ward98a,ward98b}, with values larger than 0.7 being representative
for regions located at the border between North American and Pacific plates
and small values indicating small strain regions, typical for central USA
and northwest Europe \citep{ward98a}. In addition, the cumulative moment
release is proportional to the shear modulus, that for PREM varies between
26.6 GPa at the surface to 68.2 GPa at the base of the crust
\citep{dziewonski81}.

It is important to note that the models of \citet{plesa18} and those presented
in Fig.~\ref{fig:SeisMom} do not include the contribution of stresses produced
by lithospheric flexure due to loading. Tensile-compressive stresses as
well as shear stresses distribution in the martian lithosphere correlate
with surface structures and can affect the seismic moment distribution
in particular in areas such as Tharsis, Hellas Planitia, Argyre Planitia,
Acidalia Planitia, Arcadia Planitia, and Valles Marineris
\citep{gudkova17,batov19}. In addition, although the seismicity distribution
Fig.~\ref{fig:SeisMom} includes the contribution from convective stresses
that reflect the presence of strong mantle plumes in the interior, it does
not consider the contribution from magmatic processes that may be ongoing
on Mars. This contribution may be specifically important in areas close
to the large volcanic centers in Tharsis and Elysium, and may explain the
observed seismicity in Cerberus Fossae. Thus future models that evaluate
the present-day distribution of seismicity need to include these additional
contributions.

Tectonic faults at the surface are important indicators of the internal
stress distribution in the lithosphere
\citep{banerdt92,carr74,golombek10,wise79}. However, the distribution of
present-day seismicity cannot be robustly estimated from the distribution
of these features, as this would require knowledge of which faults are
active today. The distribution of seismically active zones on present-day
Mars could range from a nearly homogeneous one, if all faults are considered
to be seismically active, to limited areas on the northern hemisphere and
in Tharsis and Elysium, if only faults cutting Amazonian terrains are active
today \citep{knapmeyer06,plesa18}. Thus, the localization of seismic events
is essential to constrain the distribution of seismicity. Maps of the location
of seismic events that were recorded by InSight could be used in future
studies to discriminate between scenarios of seismicity distribution proposed
by the analysis of various stress contributions and fault locations on
the martian surface.

\section{Conclusions and future work}
\label{sec:Conclusions}

The large amount of data and the diversity of data sets that are now available
for Mars provide a unique opportunity to investigate the planet's thermal
evolution and constrain poorly known parameters such as the mantle viscosity,
thermal variations in the interior, or the distribution of heat producing
elements between the mantle and the crust. In particular, recent results
of the InSight mission provide the most direct constraints for the martian
interior. Crustal thickness values, the size of the core and information
about the thickness of the thermal lithosphere can be combined with interior
evolution models to constrain the thermal history and present-day state
of the martian mantle and core.

Global geodynamic models show that the crustal thickness variations control
the surface heat flow and elastic lithosphere thickness pattern. The present-day
elastic lithosphere thickness at the north pole of Mars is one of the strongest
constraints for thermal evolution models and indicates that the mantle
contains less than 45\% of the total heat production and/or that the polar
cap has not yet reached elastic equilibrium.

Lithospheric temperatures show strong variations that correlate with the
crustal thickness pattern. These variations lead to a seismic velocities
pattern that can extend to depths of 400 km and deeper, depending on the
exact crustal thickness model and crustal enrichment in HPEs. The seismic
velocities variations due to temperature are larger than due to different
compositional models for the mantle.

Thermal evolution models with a large core and a dry mantle viscosity can
match the observed tidal deformation values. The size of the core indicates
that the convection pattern in the mantle is characterized by several mantle
plumes and downwellings, with stronger plumes preferentially focused in
Tharsis due to a thick insulating crust at this location. Thus, the formation
of the martian crustal dichotomy cannot be explained by an endogenous process
that would require a low degree convection pattern, but is likely the result
of the combined effects of large-scale impacts during the early martian
history and subsequent interior dynamics.

The seismicity obtained from thermal evolution models that employ the latest
crustal thickness and core radius estimates is compatible with the seismicity
derived from InSight's observations. The seismogenic layer thickness is
sensitive to the crustal thickness variations and can be used to exclude
thermal evolution models, if the depth of the events is known and
exceeds the model predicted depth.

Several open questions, however, remain and require more modeling work
and future observations. One of the largest unknowns that still remains
is the surface heat flow of Mars. While estimates of the elastic lithosphere
thickness at the north and south poles can be used to constrain the heat
loss and the thermal state of the lithosphere at those locations, future
models need to investigate scenarios, in which recent magmatic activity
in Tharsis and Elysium is compatible with a large elastic thickness at
the north pole of Mars.

Since the martian core is most likely liquid at present day, geodynamic
thermal evolution models need to investigate whether a thermally driven
dynamo can be reconciled with the duration of the martian magnetic field.
While in a recent study by \citet{greenwood21} 1D thermal evolution models
were found compatible with an internally generated magnetic field between
4.1 and 3.6 Gyr ago, it remains to be tested whether these scenarios are
compatible with additional constraints on mantle cooling imposed by the
large elastic lithosphere thickness at the north pole
\citep{broquet20} and by recent volcanic activity in Tharsis and Elysium
\citep{neukum04,vaucher09,hauber11}.

Future models need to consider constraints from the Chandler Wobble, the
movement of the pole away from the planet's average rotation axis, that
have been determined from radio tracking observations of Mars Odyssey,
Mars Reconnaissance Orbiter, and Mars Global Surveyor
\citep{konopliv20}. Since the Chandler Wobble period is sensitive to the
rheology of the martian mantle, this would provide, in addition to the
tidal quality factor Q, a valuable constraint for the mantle viscosity
and thermal state of the interior.

Seismic velocities obtained from global geodynamic models need to be combined
with seismic waves propagation to test the effect of variations in the
mantle and lithosphere on estimated travel times. While in the convective
mantle the variations in seismic velocities are typically small, temperature
variations in the lithosphere may significantly affect the travel times
of seismic waves, in particular for travel paths located on the northern
and southern hemispheres.

So far, the distribution of seismicity included only the contributions
from cooling and convective stresses. Future models need to include stresses
associated with topographic loads and evaluate the contribution from magmatic
processes. Their predictions for the distribution of seismicity could be
constrained with the observed locations of marsquakes.

\section*{Acknowledgements}
A.-C. Plesa and M. Walterova gratefully acknowledge the financial support and endorsement from the DLR Management Board Young Research Group Leader Program and the Executive Board Member for Space Research and Technology. Numerical simulations presented in this work were performed on the HoreKa supercomputer funded by the Ministry of Science, Research and the Arts Baden-W\"{u}rttemberg and by the Federal Ministry of Education and Research. M. Wieczorek acknowledges the French Space Agency (CNES) and the French National Research Agency (ANR-19-CE31-0008-08) for funding the InSight science analysis. This is InSight Publication No. 307.%

\bibliographystyle{apalike}
\bibliography{Plesa_etal}

\begin{thebibliography}{}

\bibitem[Anderson et~al., 1976]{anderson76}
Anderson, D.~L., Duennebier, F.~K., Latham, G.~V., Toks{\"o}z, M.~F., Kovach,
  R.~L., Knight, T.~C.~D., Lazarewicz, A.~R., Miller, W.~F., Nakamura, Y., and
  Sutton, G. (1976).
\newblock {The Viking Seismic Experiment}.
\newblock {\em Science}, 194(4271):1318--1321.

\bibitem[Anderson et~al., 1977]{anderson77}
Anderson, D.~L., Miller, W.~F., Latham, G.~V., Nakamura, Y., Toks{\"o}z, M.~N.,
  Dainty, A.~M., Duennebier, F.~K., Lazarewicz, A.~R., Kovach, R.~L., and
  Knight, T. C.~D. (1977).
\newblock {Seismology on Mars}.
\newblock {\em J.~Geophys.~Res.}, 82(28):4524--4546.

\bibitem[Andrews-Hanna et~al., 2008]{andrews-hanna08}
Andrews-Hanna, J.~C., Zuber, M.~T., and Banerdt, W.~B. (2008).
\newblock {The Borealis basin and the origin of the Martian crustal dichotomy}.
\newblock {\em Nature}, 453:1212--1215.

\bibitem[Bagheri et~al., 2022]{bagheri22}
Bagheri, A., Efroimsky, M., Castillo-Rogez, J., Goossens, S., Plesa, A.-C.,
  Rambaux, N., Walterov\'{a}, M., Khan, A., and Giardini, D. (2022).
\newblock {Tidal insights into the rocky and icy bodies: Review and
  perspective}.
\newblock In {Schmelzbach}, C., editor, {\em Advances in Geophysics}, volume
  63(2).

\bibitem[Bagheri et~al., 2019]{bagheri19}
Bagheri, A., Khan, A., Al-Attar, D., Crawford, O., and Giardini, D. (2019).
\newblock {Tidal response of Mars constrained from laboratory-based
  viscoelastic dissipation models and geophysical data}.
\newblock {\em Journal of Geophysical Research: Planets}, 124(11):2703--2727.

\bibitem[{Banerdt} et~al., 1992]{banerdt92}
{Banerdt}, W.~B., {Golombek}, M.~P., and {Tanaka}, K.~L. (1992).
\newblock {Stress and tectonics on Mars}.
\newblock In {George}, M., editor, {\em Mars}, pages 249--297.

\bibitem[Banerdt et~al., 2020]{banerdt20}
Banerdt, W.~B., Smrekar, S.~E., Banfield, D., Giardini, D., Golombek, M.,
  Johnson, C.~L., Lognonn{\'e}, P., Spiga, A., Spohn, T., Perrin, C., et~al.
  (2020).
\newblock {Initial results from the InSight mission on Mars}.
\newblock {\em Nature Geoscience}, pages 1--7.

\bibitem[Batov et~al., 2018]{batov19}
Batov, A., Gudkova, T., and Zharkov, V. (2018).
\newblock {Model Estimates of Non-Hydrostatic Stresses in the Martian Crust and
  Mantle: 2- Three-Level Model}.
\newblock {\em Solar System Research}, 52(3):234--240.

\bibitem[Belleguic et~al., 2005]{belleguic05}
Belleguic, V., Lognonn{\'e}, P., and Wieczorek, M.~J. (2005).
\newblock {Strength of Faults on Mars from MOLA Topography }.
\newblock {\em J.~Geophys.~Res.}, 1110:E11005.

\bibitem[Bergman, 1986]{bergman86}
Bergman, E.~A. (1986).
\newblock {Intraplate earthquakes and the state of stress in oceanic
  lithosphere}.
\newblock {\em Tectonophysics}, 132(1):1 -- 35.

\bibitem[Breuer and Moore, 2015]{breuer15}
Breuer, D. and Moore, W.~B. (2015).
\newblock {Dynamics and Thermal History of the Terrestrial Planets, the Moon,
  and Io}.
\newblock {\em Treatise on Geophysics}, 10(Second Ed.):299--348.

\bibitem[Breuer et~al., 2016]{breuer16}
Breuer, D., Plesa, A.-C., Tosi, N., and Grott, M. (2016).
\newblock {Water in the Martian interior -- The geodynamical perspective}.
\newblock {\em Meteoritic and Planetary Science}, 51(11):1959--1992.

\bibitem[Breuer et~al., 2015]{breuer15iron}
Breuer, D., Rueckriemen, T., and Spohn, T. (2015).
\newblock {Iron snow, crystal floats, and inner-core growth: modes of core
  solidification and implications for dynamos in terrestrial planets and
  moons}.
\newblock {\em Progress in Earth and Planetary Science}, 2(1):1--26.

\bibitem[Breuer and Spohn, 2003]{breuer03}
Breuer, D. and Spohn, T. (2003).
\newblock {Early plate tectonics versus single plate tectonics on Mars:
  Evidence from the magnetic field history and crust evolution}.
\newblock {\em J. Geophys. Res. E}, 108.

\bibitem[Breuer and Spohn, 2006]{breuer06}
Breuer, D. and Spohn, T. (2006).
\newblock {Viscosity of the martian mantle and its initial temperature:
  Constraints from crustal formation history and the evolution of the magnetic
  field}.
\newblock {\em Planetary and Space Science}, 54:153--169.

\bibitem[Breuer et~al., 1998]{breuer98}
Breuer, D., Yuen, D.~A., Spohn, T., and Zhang, S. (1998).
\newblock {Three dimensional models of Martian mantle convection with phase
  transitions}.
\newblock {\em Geophysical Research Letters}, 25(3):229--232.

\bibitem[Brinkman et~al., 2021]{brinkman21}
Brinkman, N., St{\"a}hler, S.~C., Giardini, D., Schmelzbach, C., Khan, A.,
  Jacob, A., Fuji, N., Perrin, C., Lognonn{\'e}, P., Beucler, E., et~al.
  (2021).
\newblock {First focal mechanisms of marsquakes}.
\newblock {\em Journal of Geophysical Research: Planets}, page e2020JE006546.

\bibitem[Broquet et~al., 2021]{broquet21}
Broquet, A., Wieczorek, M., and Fa, W. (2021).
\newblock {The composition of the south polar cap of Mars derived from orbital
  data}.
\newblock {\em Journal of Geophysical Research: Planets}, 126(8):e2020JE006730.

\bibitem[Broquet et~al., 2020]{broquet20}
Broquet, A., Wieczorek, M.~A., and Fa, W. (2020).
\newblock {Flexure of the lithosphere beneath the North Polar Cap of Mars:
  Implications for ice composition and heat flow}.
\newblock {\em Geophysical Research Letters}, 47(5):e2019GL086746.

\bibitem[Burov and Diament, 1995]{burov95}
Burov, E.-B. and Diament, M. (1995).
\newblock {The effective elastic thickness (Te) of continental lithosphere:
  What does it really mean?}
\newblock {\em J.~Geophys.~Res.}, 100:3905--3927.

\bibitem[Carr, 1974]{carr74}
Carr, M.~H. (1974).
\newblock {Tectonism and volcanism of the Tharsis region of Mars}.
\newblock {\em J.~Geophys.~Res.}, 79(26):3943--3949.

\bibitem[Castillo-Rogez et~al., 2011]{castillo-rogez11}
Castillo-Rogez, J.~C., Efroimsky, M., and Lainey, V. (2011).
\newblock {The tidal history of Iapetus: Spin dynamics in the light of a
  refined dissipation model}.
\newblock {\em Journal of Geophysical Research: Planets}, 116(E9).

\bibitem[Christensen and Yuen, 1985]{christensen85}
Christensen, U.~R. and Yuen, D.~A. (1985).
\newblock {Layered convection induced by phase transitions}.
\newblock {\em J.~Geophys.~Res.}, 90:10291--10300.

\bibitem[Collinet et~al., 2015]{collinet15}
Collinet, M., M{\'e}dard, E., Charlier, B., Vander~Auwera, J., and Grove, T.~L.
  (2015).
\newblock {Melting of the primitive Martian mantle at 0.5--2.2 GPa and the
  origin of basalts and alkaline rocks on Mars}.
\newblock {\em Earth and Planetary Science Letters}, 427:83--94.

\bibitem[Connolly, 2009]{connolly09}
Connolly, J. (2009).
\newblock {The geodynamic equation of state: what and how}.
\newblock {\em Geochemistry, Geophysics, Geosystems}, 10(10).

\bibitem[Davies and Pommier, 2018]{davies18}
Davies, C.~J. and Pommier, A. (2018).
\newblock {Iron snow in the Martian core?}
\newblock {\em Earth and Planetary Science Letters}, 481:189--200.

\bibitem[Dziewonski and Anderson, 1981]{dziewonski81}
Dziewonski, A.~M. and Anderson, D.~L. (1981).
\newblock {Preliminary reference Earth model}.
\newblock {\em Physics of the earth and planetary interiors}, 25(4):297--356.

\bibitem[Efroimsky, 2012]{efroimsky12}
Efroimsky, M. (2012).
\newblock {Tidal dissipation compared to seismic dissipation: In small bodies,
  Earths, and super-Earths}.
\newblock {\em The Astrophysical Journal}, 746(2):150.

\bibitem[Ewing et~al., 1971]{ewing71}
Ewing, M., Latham, G., Press, F., Sutton, G., Dorman, J., Nakamura, Y.,
  Meissner, R., Duennebier, F., and Kovach, R. (1971).
\newblock {Seismology of the Moon and implications on internal structure,
  origin and evolution}.
\newblock {\em Highlights of astronomy}, 2:155--172.

\bibitem[Filiberto and Dasgupta, 2015]{filiberto15}
Filiberto, J. and Dasgupta, R. (2015).
\newblock {Constraints on the depth and thermal vigor of melting in the Martian
  mantle}.
\newblock {\em Journal of Geophysical Research: Planets}, 120(1):109--122.

\bibitem[Folkner et~al., 2018]{folkner18}
Folkner, W.~M., Dehant, V., Le~Maistre, S., Yseboodt, M., Rivoldini, A.,
  Van~Hoolst, T., Asmar, S.~W., and Golombek, M.~P. (2018).
\newblock {The rotation and interior structure experiment on the InSight
  mission to Mars}.
\newblock {\em Space Science Reviews}, 214(5):100.

\bibitem[Fraeman and Korenaga, 2010]{fraeman10}
Fraeman, A.~A. and Korenaga, J. (2010).
\newblock {The influence of mantle melting on the evolution of Mars}.
\newblock {\em Icarus}, 210(1):43--57.

\bibitem[Frey and Schultz, 1988]{frey88}
Frey, H. and Schultz, R.~A. (1988).
\newblock {Large impact basins and the mega-impact origin for the crustal
  dichotomy on Mars}.
\newblock {\em Geophysical Research Letters}, 15:229--232.

\bibitem[Genova et~al., 2016]{genova16}
Genova, A., Goossens, S., Lemoine, F.~G., Mazarico, E., Neumann, G.~A., Smith,
  D.~E., and Zuber, M.~T. (2016).
\newblock {Seasonal and static gravity field of Mars from MGS, Mars Odyssey and
  MRO radio science}.
\newblock {\em Icarus}, 272(Supplement C):228 -- 245.

\bibitem[Giardini et~al., 2020]{giardini20}
Giardini, D., Lognonn{\'e}, P., Banerdt, W.~B., Pike, W.~T., Christensen, U.,
  Ceylan, S., Clinton, J.~F., van Driel, M., St{\"a}hler, S.~C., B{\"o}se, M.,
  et~al. (2020).
\newblock {The seismicity of Mars}.
\newblock {\em Nature Geoscience}, 13(3):205--212.

\bibitem[Golabek et~al., 2011]{golabek11}
Golabek, G., Keller, T., Gerya, T.~V., Zhu, G., Tackley, P.~J., and Connolly,
  J.~A.~D. (2011).
\newblock {Origin of the martian dichotomy and Tharsis from a giant impact
  causing massive magmatism}.
\newblock {\em Icarus}, 215:346--357.

\bibitem[Golombek, 1994]{golombek94}
Golombek, M.~P. (1994).
\newblock {Constraints on the Largest Marsquake}.
\newblock {\it LPSC} XXV, Houston, Texas.
\newblock {Abstract \#1221}.

\bibitem[Golombek, 2002]{golombek02}
Golombek, M.~P. (2002).
\newblock {A revision of Mars seismicity from surface faulting}.
\newblock {\it LPS} XXXIII, Houston, Texas.
\newblock {Abstract \#1244}.

\bibitem[Golombek et~al., 1992]{golombek92}
Golombek, M.~P., Banerdt, W.~B., Tanaka, K.~L., and Tralli, D.~M. (1992).
\newblock {A Prediction of Mars Seismicity from Surface Faulting}.
\newblock {\em Science}, 258(5084):979--981.

\bibitem[Golombek and Phillips, 2010]{golombek10}
Golombek, M.~P. and Phillips, R.~J. (2010).
\newblock {Mars Tectonics}.
\newblock In Watters, T.~R. and Schultz, R.~A., editors, {\em {Planetary
  Tectonics}}, pages 183--232. Cambridge University Press.

\bibitem[Greeley and Schneid, 1991]{greeley96}
Greeley, R. and Schneid, B.~D. (1991).
\newblock {Magma Generation on Mars: Amounts, Rates, and Comparisons with
  Earth, Moon, and Venus}.
\newblock {\em Science}, 254(5034):996--998.

\bibitem[Greenwood et~al., 2021]{greenwood21}
Greenwood, S., Davies, C.~J., and Pommier, A. (2021).
\newblock {Influence of thermal stratification on the structure and evolution
  of the Martian core}.
\newblock {\em Geophysical Research Letters}, 48(22):e2021GL095198.

\bibitem[Grott et~al., 2013]{grott13}
Grott, M., Baratoux, D., Hauber, E., Sautter, V., Mustard, J., Gasnault, O.,
  Ruff, S.~W., Karato, S.-I., Debaille, V., Knapmeyer, M., Sohl, F., Hoolst,
  T.~V., Breuer, D., Morschhauser, A., and Toplis, M.~J. (2013).
\newblock {Long-Term Evolution of the Martian Crust-Mantle System}.
\newblock {\em Space Science Review}, 172(1):49--111.

\bibitem[Grott and Breuer, 2008]{grott08}
Grott, M. and Breuer, D. (2008).
\newblock {The evolution of the Martian elastic lithosphere and implications
  for crustal and mantle rheology}.
\newblock {\em Icarus}, 193:503--515.

\bibitem[Grott and Breuer, 2010]{grott10}
Grott, M. and Breuer, D. (2010).
\newblock {On the spatial variability of the Martian elastic lithosphere
  thickness: Evidence for mantle plumes?}
\newblock {\em J.~Geophys.~Res.}, 115(E3).

\bibitem[Grott et~al., 2011]{grott11}
Grott, M., Morschhauser, A., Breuer, D., and Hauber, E. (2011).
\newblock {Volcanic outgassing of CO$_2$ and H$_2$O on Mars}.
\newblock {\em Earth and Planetary Science Letters}, 308:391--400.

\bibitem[Gudkova et~al., 2017]{gudkova17}
Gudkova, T., Batov, A., and Zharkov, V. (2017).
\newblock {Model estimates of non-hydrostatic stresses in the Martian crust and
  mantle: 1—Two-level model}.
\newblock {\em Solar System Research}, 51(6):457--478.

\bibitem[Gudkova and Zharkov, 2004]{gudkova04}
Gudkova, T. and Zharkov, V. (2004).
\newblock {Mars: interior structure and excitation of free oscillations}.
\newblock {\em Physics of the Earth and Planetary Interiors}, 142(1-2):1--22.

\bibitem[Hahn et~al., 2011]{hahn11}
Hahn, B.~C., McLennan, S.~M., and Klein, E.~C. (2011).
\newblock {Martian surface heat production and crustal heat flow from Mars
  Odyssey Gamma-Ray spectrometry}.
\newblock {\em Geophysical Research Letters}, 38(14).

\bibitem[Harder and Christensen, 1996]{harder96}
Harder, H. and Christensen, U. (1996).
\newblock A one-plume model of martian mantle convection.
\newblock {\em Nature}, 380:507--509.

\bibitem[Hauber et~al., 2011]{hauber11}
Hauber, E., Bro{\v{z}}, P., Jagert, F., Jod{\l}owski, P., and Platz, T. (2011).
\newblock {Very recent and wide-spread basaltic volcanism on Mars}.
\newblock {\em Geophysical Research Letters}, 38(L10201).

\bibitem[Hauck and Phillips, 2002]{hauck02}
Hauck, S.~A. and Phillips, R.~P. (2002).
\newblock {Thermal and crustal evolution of Mars}.
\newblock {\em J.~Geophys.~Res.}, 107(E7):5052.

\bibitem[Heister et~al., 2017]{heister17}
Heister, T., Dannberg, J., Gassm{\"o}ller, R., and Bangerth, W. (2017).
\newblock {High accuracy mantle convection simulation through modern numerical
  methods--II: realistic models and problems}.
\newblock {\em Geophysical Journal International}, 210(2):833--851.

\bibitem[Helffrich, 2017]{helffrich17}
Helffrich, G. (2017).
\newblock {Mars core structure—concise review and anticipated insights from
  InSight}.
\newblock {\em Progress in Earth and Planetary Science}, 4(1):1--14.

\bibitem[Hemingway and Driscoll, 2021]{hemingway21}
Hemingway, D.~J. and Driscoll, P.~E. (2021).
\newblock {History and future of the Martian dynamo and implications of a
  hypothetical solid inner core}.
\newblock {\em Journal of Geophysical Research: Planets}, 126(4):e2020JE006663.

\bibitem[Hirth and Kohlstedt, 2003]{hirth13}
Hirth, G. and Kohlstedt, D. (2003).
\newblock {\em {Rheology of the Upper Mantle and the Mantle Wedge: A View from
  the Experimentalists}}, pages 83--105.
\newblock American Geophysical Union.

\bibitem[Horleston et~al., 2022]{horleston22}
Horleston, A.~C., Clinton, J.~F., Ceylan, S., Giardini, D., Charalambous, C.,
  Irving, J. C.~E., Lognonné, P., Stähler, S.~C., Zenhäusern, G., Dahmen,
  N.~L., Duran, C., Kawamura, T., Khan, A., Kim, D., Plasman, M., Euchner, F.,
  Beghein, C., Beucler, E., Huang, Q., Knapmeyer, M., Knapmeyer‐Endrun, B.,
  Lekić, V., Li, J., Perrin, C., Schimmel, M., Schmerr, N.~C., Stott, A.~E.,
  Stutzmann, E., Teanby, N.~A., Xu, Z., Panning, M., and Banerdt, W.~B. (2022).
\newblock {The Far Side of Mars: Two Distant Marsquakes Detected by InSight}.
\newblock {\em The Seismic Record}, 2(2):88--99.

\bibitem[{InSight Marsquake Service}, 2022]{mqs22}
{InSight Marsquake Service} (2022).
\newblock {Mars Seismic Catalogue, InSight Mission; V9 2022-01-01. ETHZ, IPGP,
  JPL, ICL, Univ. Bristol}.

\bibitem[Kagan, 2002]{kagan02}
Kagan, Y.~Y. (2002).
\newblock {Seismic moment distribution revisited: II. Moment conservation
  principle}.
\newblock {\em Geophys.~J.~Int.}, 149:731--754.

\bibitem[Karato and Wu, 1993]{karato93}
Karato, S.~I. and Wu, P. (1993).
\newblock Rheology of the upper mantle: a synthesis.
\newblock {\em Science}, 260:771--778.

\bibitem[{Kaula} et~al., 1981]{bvsp81}
{Kaula}, W.~M., {Head}, J.~W., I., {Merrill}, R.~B., {Pepin}, R.~O., {Solomon},
  S.~C., {Walker}, D., and {Wood}, C.~A. (1981).
\newblock {\em {Basaltic volcanism on the terrestrial planets.}}

\bibitem[Kedar et~al., 2021]{kedar21}
Kedar, S., Panning, M.~P., Smrekar, S.~E., St{\"a}hler, S.~C., King, S.~D.,
  Golombek, M.~P., Manga, M., Julian, B.~R., Shiro, B., Perrin, C., Power, J.,
  Michaut, C., Ceylan, S., Giardini, D., P.Lognonn{\'e}, and Bandert, W.~B.
  (2021).
\newblock {Analyzing Low Frequency Seismic Events at Cerberus Fossae as Long
  Period Volcanic Quakes}.
\newblock {\em J.~Geophys.~Res.}

\bibitem[Keller and Tackley, 2009]{keller09}
Keller, T. and Tackley, P.~J. (2009).
\newblock {Towards self-consistent modeling of the martian dichotomy: The
  influence of one-ridge convection on crustal thickness distribution}.
\newblock {\em Icarus}, 202:429--443.

\bibitem[Khan et~al., 2021]{khan21}
Khan, A., Ceylan, S., van Driel, M., Giardini, D., Lognonn{\'e}, P., Samuel,
  H., Schmerr, N.~C., St{\"a}hler, S.~C., Duran, A.~C., Huang, Q., et~al.
  (2021).
\newblock {Upper mantle structure of Mars from InSight seismic data}.
\newblock {\em Science}, 373(6553):434--438.

\bibitem[Khan et~al., 2018]{khan17}
Khan, A., Liebske, C., Rozel, A., Rivoldini, A., Nimmo, F., Connolly, J. A.~D.,
  Plesa, A., and Giardini, D. (2018).
\newblock {A Geophysical Perspective on the Bulk Composition of Mars}.
\newblock {\em J.~Geophys.~Res.}, 123(2):575--611.

\bibitem[Khan et~al., 2022]{khan22}
Khan, A., Sossi, P., Liebske, C., Rivoldini, A., and Giardini, D. (2022).
\newblock {Geophysical and cosmochemical evidence for a volatile-rich Mars}.
\newblock {\em Earth and Planetary Science Letters}, 578:117330.

\bibitem[Kiefer et~al., 2015]{kiefer15}
Kiefer, W.~S., Filiberto, J., Sandu, C., and Li, Q. (2015).
\newblock {The effects of mantle composition on the peridotite solidus:
  Implications for the magmatic history of Mars}.
\newblock {\em Geochimica et Cosmochimica Acta}, 162:247--258.

\bibitem[Kiefer and Li, 2009]{kiefer09}
Kiefer, W.~S. and Li, Q. (2009).
\newblock {Mantle convection controls the observed lateral variations in
  lithospheric thickness on present-day Mars}.
\newblock {\em Geophysical Research Letters}, 36:L18203.

\bibitem[Kiefer and Li, 2016]{kiefer16}
Kiefer, W.~S. and Li, Q. (2016).
\newblock {Water undersaturated mantle plume volcanism on present-day Mars}.
\newblock {\em Meteoritics \& Planetary Science}, 51(11):1993--2010.

\bibitem[Kim et~al., 2021]{kim21}
Kim, D., Leki{\'c}, V., Irving, J., Schmerr, N., Knapmeyer-Endrun, B., Joshi,
  R., Panning, M., Tauzin, B., Karakostas, F., Maguire, R., et~al. (2021).
\newblock {Improving constraints on planetary interiors with PPS receiver
  functions}.
\newblock {\em Journal of Geophysical Research: Planets},
  126(11):e2021JE006983.

\bibitem[Knapmeyer et~al., 2006]{knapmeyer06}
Knapmeyer, M., Oberst, J., Hauber, E., W{\"a}hlisch, M., Deuchler, C., and
  Wagner, R. (2006).
\newblock Working models for spatial distribution and level of mars'
  seismicity.
\newblock {\em JGR}, 111(E11).

\bibitem[Knapmeyer et~al., 2021]{knapmeyer21}
Knapmeyer, M., St{\"a}hler, S.~C., Daubar, I., Forget, F., Spiga, A., Pierron,
  T., van Driel, M., Banfield, D., Hauber, E., Grott, M., et~al. (2021).
\newblock {Seasonal seismic activity on Mars}.
\newblock {\em Earth and Planetary Science Letters}, 576:117171.

\bibitem[Knapmeyer and Walterov\'{a}, 2022]{knapmeyer22}
Knapmeyer, M. and Walterov\'{a}, M. (2022).
\newblock {Planetary Core Radii: From Plato towards PLATO}.
\newblock In {Schmelzbach}, C., editor, {\em Advances in Geophysics}, volume
  63(2).

\bibitem[Knapmeyer-Endrun et~al., 2021]{knapmeyer-endrun21}
Knapmeyer-Endrun, B., Panning, M.~P., Bissig, F., Joshi, R., Khan, A., Kim, D.,
  Leki{\'c}, V., Tauzin, B., Tharimena, S., Plasman, M., et~al. (2021).
\newblock {Thickness and structure of the martian crust from InSight seismic
  data}.
\newblock {\em Science}, 373(6553):438--443.

\bibitem[Konopliv et~al., 2016]{konopliv16}
Konopliv, A.~S., Park, R.~S., and Folkner, W.~M. (2016).
\newblock {An improved JPL Mars gravity field and orientation from Mars orbiter
  and lander tracking data}.
\newblock {\em Icarus}, 274(Supplement C):253 -- 260.

\bibitem[Konopliv et~al., 2020]{konopliv20}
Konopliv, A.~S., Park, R.~S., Rivoldini, A., Baland, R.-M., Le~Maistre, S.,
  Van~Hoolst, T., Yseboodt, M., and Dehant, V. (2020).
\newblock {Detection of the Chandler wobble of Mars from orbiting spacecraft}.
\newblock {\em Geophysical Research Letters}, 47(21):e2020GL090568.

\bibitem[Kronbichler et~al., 2012]{kronbichler12}
Kronbichler, M., Heister, T., and Bangerth, W. (2012).
\newblock {High accuracy mantle convection simulation through modern numerical
  methods}.
\newblock {\em Geophysical Journal International}, 191(1):12--29.

\bibitem[Lainey, 2016]{lainey16}
Lainey, V. (2016).
\newblock {Quantification of tidal parameters from Solar System data}.
\newblock {\em {Celest. Mech. Dyn. Astr.}}, 126:145--156.

\bibitem[Lawrence et~al., 2000]{lawrence00}
Lawrence, D.~J., Feldman, W.~C., Barraclough, B.~L., Binder, A.~B., Elphic,
  R.~C., Maurice, S., Miller, M.~C., and Prettyman, T.~H. (2000).
\newblock {Thorium abundances on the lunar surface}.
\newblock {\em J.~Geophys.~Res.}, 105(E8):20307--20331.

\bibitem[Lognonn{\'e} et~al., 2020]{lognonne20}
Lognonn{\'e}, P., Banerdt, W., Pike, W., Giardini, D., Christensen, U., Garcia,
  R., Kawamura, T., Kedar, S., Knapmeyer-Endrun, B., Margerin, L., et~al.
  (2020).
\newblock {Constraints on the shallow elastic and anelastic structure of Mars
  from InSight seismic data}.
\newblock {\em Nature Geoscience}, 13(3):213--220.

\bibitem[Marinova et~al., 2008]{marinova08}
Marinova, M.~M., Aharonson, O., and Asphaug, E. (2008).
\newblock Mega-impact formation of the mars hemispheric dichotomy.
\newblock {\em Nature}, 453:1216--1219.

\bibitem[Matsukage et~al., 2013]{matsukage13}
Matsukage, K.~N., Nagayo, Y., Whitaker, M.~L., Takahashi, E., and Kawasaki, T.
  (2013).
\newblock {Melting of the Martian mantle from 1.0 to 4.5 GPa}.
\newblock {\em Journal of Mineralogical and Petrological Sciences}, page
  120820.

\bibitem[McCubbin et~al., 2016]{mcCubbin16}
McCubbin, F.~M., Boyce, J.~W., Srinivasan, P., Santos, A.~R., Elardo, S.~M.,
  Filiberto, J., Steele, A., and Shearer, C.~K. (2016).
\newblock {Heterogeneous distribution of H2O in the Martian interior:
  Implications for the abundance of H2O in depleted and enriched mantle
  sources}.
\newblock {\em Meteoritics \& Planetary Science}, 51(11):2036--2060.

\bibitem[McGovern et~al., 2004]{mcgovern04}
McGovern, P.~J., Solomon, S.~C., Smith, D.~E., Zuber, M.~T., Simons, M.,
  Wieczorek, M.~A., Phillips, R.~J., Neumann, G.~A., Aharonson, O., and Head,
  J.~W. (2004).
\newblock {Correction to “Localized gravity/topography admittance and
  correlation spectra on Mars: Implications for regional and global
  evolution”}.
\newblock {\em J.~Geophys.~Res.}, 109(E7).

\bibitem[McNutt, 1984]{mcnutt84}
McNutt, M.~K. (1984).
\newblock {Lithospheric flexure and thermal anomalies}.
\newblock {\em {Journal of Geophysical Research: Solid Earth}},
  89(B13):11180--11194.

\bibitem[Michel and Forni, 2011]{michel11}
Michel, N. and Forni, O. (2011).
\newblock {Mars mantle convection: Influence of phase transitions with core
  cooling}.
\newblock {\em Planetary and Space Science}, 59:741--748.

\bibitem[Mittelholz et~al., 2020]{mittelholz20}
Mittelholz, A., Johnson, C., Feinberg, J., Langlais, B., and Phillips, R.
  (2020).
\newblock {Timing of the martian dynamo: New constraints for a core field 4.5
  and 3.7 Ga ago}.
\newblock {\em Science advances}, 6(18):eaba0513.

\bibitem[Mocquet and Menvielle, 2000]{mocquet00}
Mocquet, A. and Menvielle, M. (2000).
\newblock {Complementarity of seismological and electromagnetic sounding
  methods for constraining the structure of the Martian mantle}.
\newblock {\em Planetary and Space Science}, 48(12-14):1249--1260.

\bibitem[Mocquet et~al., 1996]{mocquet96}
Mocquet, A., Vacher, P., Grasset, O., and Sotin, C. (1996).
\newblock {Theoretical seismic models of Mars: the importance of the iron
  content of the mantle}.
\newblock {\em Planetary and space science}, 44(11):1251--1268.

\bibitem[Mori et~al., 2017]{mori17}
Mori, Y., Ozawa, H., Hirose, K., Sinmyo, R., Tateno, S., Morard, G., and
  Ohishi, Y. (2017).
\newblock {Melting experiments on Fe--Fe3S system to 254 GPa}.
\newblock {\em Earth and Planetary Science Letters}, 464:135--141.

\bibitem[Morschhauser et~al., 2011]{morschhauser11}
Morschhauser, A., Grott, M., and Breuer, D. (2011).
\newblock Crustal recycling, mantle dehydration, and the thermal evolution of
  {Mars}.
\newblock {\em Icarus}, 212:541--558.

\bibitem[Neukum et~al., 2004]{neukum04}
Neukum, G., Jaumann, R., Hoffmann, H., Hauber, E., Head, J., Basilevsky, A.,
  Ivanov, B.~A., Werner, S.~C., van Gasselt, S., Murray, J.~B., McCord, T., and
  Team, T. H. C.-I. (2004).
\newblock {Recent and episodic volcanic and glacial activity on Mars revealed
  by the High Resolution Stereo Camera}.
\newblock {\em Nature}, 432:971--979.

\bibitem[Neumann et~al., 2004]{neumann04}
Neumann, G.~A., Zuber, M.~T., Wieczorek, M.~A., McGovern, P.~J., Lemoine,
  F.~G., and Smith, D.~E. (2004).
\newblock {Crustal structure of Mars from gravity and topography}.
\newblock {\em J.~Geophys.~Res.}, 109:E08002.

\bibitem[Nimmo and Faul, 2013]{nimmo13}
Nimmo, F. and Faul, U.~H. (2013).
\newblock {Dissipation at tidal and seismic frequencies in a melt-free,
  anhydrous Mars}.
\newblock {\em J.~Geophys.~Res.}, 118(12):2558--2569.

\bibitem[Nimmo et~al., 2008]{nimmo08}
Nimmo, F., Hart, S.~D., Korycansky, D.~G., and Agnor, C.~B. (2008).
\newblock {Implications of an impact origin for the martian hemispheric
  dichotomy}.
\newblock {\em Nature}, 453:1220--1223.

\bibitem[Nimmo and Tanaka, 2005]{nimmo05}
Nimmo, F. and Tanaka, K. (2005).
\newblock {Early Crustal Evolution of Mars}.
\newblock {\em Annual Review of Earth and Planetary Sciences}, 33(1):133--161.

\bibitem[Oberst, 1987]{oberst87}
Oberst, J. (1987).
\newblock {Unusually high stress drops associated with shallow moonquakes}.
\newblock {\em J.~Geophys.~Res.}, 92:1397--1405.

\bibitem[Phillips, 1991]{phillips91}
Phillips, R.~J. (1991).
\newblock {Expected rates of Marsquakes, in Scientific Rationale and
  Requirements for a Global Seismic Network on Mars}.
\newblock {\em LPI Tech. Rep.}, 91-02 LPI/TR-91-02:35--38.
\newblock {Lunar and Planet. Inst., Houston, Texas}.

\bibitem[Phillips et~al., 2008]{phillips08}
Phillips, R.~J., Zuber, M.~T., Smrekar, S.~E., Mellon, M.~T., Head, J.~W.,
  Tanaka, K.~L., Putzig, N.~E., Milkovich, S.~M., Campbell, B.~A., Plaut,
  J.~J., Safaeinili, A., Seu, R., Biccari, D., Carter, L.~M., Picardi, G.,
  Orosei, R., Mohit, P.~S., Heggy, E., Zurek, R.~W., Egan, A.~F., Giacomoni,
  E., Russo, F., Cutigni, M., Pettinelli, E., Holt, J.~W., Leuschen, C.~J., and
  Marinangeli, L. (2008).
\newblock {Mars north polar deposits: Stratigraphy, age, and geodynamical
  response}.
\newblock {\em Science}, 320(5880):1182--1185.

\bibitem[Plesa et~al., 2021]{plesa21}
Plesa, A.-C., Bozda{\u{g}}, E., Rivoldini, A., Knapmeyer, M., McLennan, S.~M.,
  Padovan, S., Tosi, N., Breuer, D., Peter, D., Staehler, S., et~al. (2021).
\newblock {Seismic Velocity Variations in a 3D Martian Mantle: Implications for
  the InSight Measurements}.
\newblock {\em Journal of Geophysical Research: Planets}, page e2020JE006755.

\bibitem[Plesa and Breuer, 2014]{plesa14b}
Plesa, A.-C. and Breuer, D. (2014).
\newblock {Partial melting in one-plate planets: Implications for
  thermo-chemical and atmospheric evolution}.
\newblock {\em Planetary and Space Science}, 98:50 -- 65.
\newblock Planetary evolution and life.

\bibitem[Plesa et~al., 2016]{plesa16}
Plesa, A.-C., Grott, M., Tosi, N., Breuer, D., Spohn, T., and Wieczorek, M.~A.
  (2016).
\newblock {How large are present-day heat flux variations across the surface of
  Mars?}
\newblock {\em J.~Geophys.~Res.}, 121(12):2386--2403.

\bibitem[Plesa et~al., 2018a]{plesa18}
Plesa, A.-C., Knapmeyer, M., Golombek, M., Breuer, D., Grott, M., Kawamura, T.,
  Lognonn{\'e}, P., Tosi, N., and Weber, R. (2018a).
\newblock {Present-Day Mars' Seismicity Predicted From 3-D Thermal Evolution
  Models of Interior Dynamics}.
\newblock {\em Geophysical Research Letters}, 45(6):2580--2589.

\bibitem[Plesa et~al., 2018b]{plesa18b}
Plesa, A.-C., Padovan, S., Tosi, N., Breuer, D., Grott, M., Wieczorek, M.,
  Spohn, T., Smrekar, S., and Banerdt, W. (2018b).
\newblock {The thermal state and interior structure of Mars}.
\newblock {\em Geophysical Research Letters}, 45(22):12--198.

\bibitem[Ray et~al., 2001]{ray01}
Ray, R.~D., Eanes, R.~J., and Lemoine, F.~G. (2001).
\newblock {Constraints on energy dissipation in the Earth's body tide from
  satellite tracking and altimetry}.
\newblock {\em Geophysical Journal International}, 144(2):471--480.

\bibitem[Rivoldini et~al., 2011]{rivoldini11}
Rivoldini, A., Hoolst, T.~V., Verhoeven, O., Mocquet, A., and Dehant, V.
  (2011).
\newblock {Geodesy constraints on the interior structure and composition of
  Mars}.
\newblock {\em Icarus}, 213(2):451 -- 472.

\bibitem[Roberts and Arkani-Hamed, 2017]{roberts17}
Roberts, J. and Arkani-Hamed, J. (2017).
\newblock {Effects of basin-forming impacts on the thermal evolution and
  magnetic field of Mars}.
\newblock {\em Earth and Planetary Science Letters}, 478(Supplement C):192 --
  202.

\bibitem[Roberts and Zhong, 2006]{roberts06}
Roberts, J.~H. and Zhong, S. (2006).
\newblock {Degree-1 convection in the Martian mantle and the origin of the
  hemispheric dichotomy}.
\newblock {\em J.~Geophys.~Res. (Planets)}, 111.

\bibitem[Ruedas and Breuer, 2017]{ruedas17}
Ruedas, T. and Breuer, D. (2017).
\newblock {On the relative importance of thermal and chemical buoyancy in
  regular and impact-induced melting in a Mars-like planet}.
\newblock {\em Journal of Geophysical Research: Planets}, 122(7):1554--1579.
\newblock 2016JE005221.

\bibitem[Ruedas et~al., 2013]{ruedas13}
Ruedas, T., Tackley, P.~J., and Solomon, S.~C. (2013).
\newblock {Thermal and compositional evolution of the martian mantle: Effects
  of phase transitions and melting}.
\newblock {\em Physics of the Earth and Planetary Interios}, 216:32--58.

\bibitem[Sabadini and Vermeersen, 2004]{sabadini04}
Sabadini, R. and Vermeersen, B. (2004).
\newblock {Normal mode theory in viscoelasticity}.
\newblock In {\em Global Dynamics of the Earth}, pages 1--44. Springer.

\bibitem[Samuel et~al., 2021]{samuel21}
Samuel, H., Ballmer, M.~D., Padovan, S., Tosi, N., Rivoldini, A., and Plesa,
  A.-c. (2021).
\newblock {The Thermo-Chemical Evolution of Mars With a Strongly Stratified
  Mantle}.
\newblock {\em Journal of Geophysical Research: Planets}, 126(4):e2020JE006613.

\bibitem[Samuel et~al., 2019]{samuel19}
Samuel, H., Lognonn{\'e}, P., Panning, M., and Lainey, V. (2019).
\newblock {The rheology and thermal history of Mars revealed by the orbital
  evolution of Phobos}.
\newblock {\em Nature}, 569(7757):523--527.

\bibitem[Scholz, 1998]{scholz98}
Scholz, C.~H. (1998).
\newblock {Earthquakes and friction laws}.
\newblock {\em Nature}, 391:37--42.

\bibitem[Schubert et~al., 2001]{schubert01}
Schubert, G., Turcotte, D.~L., and Olson, P. (2001).
\newblock {\em {Mantle convection in the Earth and planets}}.
\newblock Cambridge University Press, Cambridge.

\bibitem[Schumacher and Breuer, 2006]{schumacher06}
Schumacher, S. and Breuer, D. (2006).
\newblock {Influence of a variable thermal conductivity on the thermochemical
  evolution of Mars}.
\newblock {\em J.~Geophys.~Res.}, 111(E02006).

\bibitem[Smrekar et~al., 2019]{smrekar19}
Smrekar, S.~E., Lognonn{\'e}, P., Spohn, T., Banerdt, W.~B., Breuer, D.,
  Christensen, U., Dehant, V., Drilleau, M., Folkner, W., Fuji, N., et~al.
  (2019).
\newblock {Pre-mission InSights on the Interior of Mars}.
\newblock {\em Space Science Reviews}, 215(1):3.

\bibitem[Sohl and Spohn, 1997]{sohl97}
Sohl, F. and Spohn, T. (1997).
\newblock {The interior structure of Mars: Implications from SNC meteorites}.
\newblock {\em J. Geophys. Res.}, 102(E1):1613--1635.

\bibitem[Spohn et~al., 2018]{spohn18}
Spohn, T., Grott, M., Smrekar, S., Knollenberg, J., Hudson, T., Krause, C.,
  M{\"u}ller, N., J{\"a}nchen, J., B{\"o}rner, A., Wippermann, T., et~al.
  (2018).
\newblock {The heat flow and physical properties package (HP$^3$) for the
  InSight mission}.
\newblock {\em Space Science Reviews}, 214(5):96.

\bibitem[Spohn et~al., 1998]{spohn98}
Spohn, T., Sohl, F., and Breuer, D. (1998).
\newblock {Mars}.
\newblock {\em Astron. Astrophys. Rev.}, 8:181--235.

\bibitem[St{\"a}hler and Knapmeyer, 2022]{staehler22}
St{\"a}hler, S. and Knapmeyer, M. (2022).
\newblock {Seismology in the Solar System}.
\newblock In {Schmelzbach}, C., editor, {\em Advances in Geophysics}, volume
  63(2).

\bibitem[St{\"a}hler et~al., 2021]{staehler21}
St{\"a}hler, S.~C., Khan, A., Banerdt, W.~B., Lognonn{\'e}, P., Giardini, D.,
  Ceylan, S., Drilleau, M., Duran, A.~C., Garcia, R.~F., Huang, Q., et~al.
  (2021).
\newblock {Seismic detection of the martian core}.
\newblock {\em Science}, 373(6553):443--448.

\bibitem[Steinbach and Yuen, 1994]{steinbach94}
Steinbach, V. and Yuen, D. (1994).
\newblock {Effects of depth-dependent properties on the thermal anomalies
  produced in flush instabilities from phase transitions}.
\newblock {\em Physics of the Earth and Planetary Interios}, 86:165--183.

\bibitem[Stevenson et~al., 1983]{stevenson83}
Stevenson, D.~J., Spohn, T., and Schubert, G. (1983).
\newblock {Magnetism and Thermal Evolution of the Terrestrial Planets}.
\newblock {\em Icarus}, 54:466--489.

\bibitem[Stewart et~al., 2007]{stewart07}
Stewart, A.~J., Schmidt, M.~W., Van~Westrenen, W., and Liebske, C. (2007).
\newblock {Mars: A new core-crystallization regime}.
\newblock {\em Science}, 316(5829):1323--1325.

\bibitem[Stixrude and Lithgow-Bertelloni, 2011]{stixrude11}
Stixrude, L. and Lithgow-Bertelloni, C. (2011).
\newblock {Thermodynamics of mantle minerals-II. Phase equilibria}.
\newblock {\em Geophysical Journal International}, 184(3):1180--1213.

\bibitem[Taylor, 2013]{taylor13}
Taylor, G.~J. (2013).
\newblock {The bulk composition of Mars}.
\newblock {\em Geochemistry}, 73(4):401--420.

\bibitem[Taylor et~al., 2006a]{taylor06}
Taylor, G.~J., Boynton, W., Br{\"u}ckner, J., W{\"a}nke, H., Dreibus, G.,
  Kerry, K., Keller, J., Reedy, R., Evans, L., Starr, R., Squyres, S.,
  Karunatillake, S., Gasnault, O., Maurice, S., {d'Uston}, C., Englert, P.,
  Dohm, J., Baker, V., Hamara, D., Janes, D., Sprague, A., Kim, K., and Drake,
  D. (2006a).
\newblock {Bulk composition and early differentiation of Mars}.
\newblock {\em J.~Geophys.~Res.}, 111(E3).

\bibitem[Taylor et~al., 2006b]{taylor06b}
Taylor, G.~J., Stopar, J., Boynton, W.~V., Karunatillake, S., Keller, J.~M.,
  Br{\"u}ckner, J., W{\"a}nke, H., Dreibus, G., Kerry, K.~E., Reedy, R.~C.,
  et~al. (2006b).
\newblock {Variations in K/Th on Mars}.
\newblock {\em Journal of Geophysical Research: Planets}, 111(E3).

\bibitem[Taylor et~al., 2013]{taylor13Cerberus}
Taylor, J., Teanby, N.~A., and Wookey, J. (2013).
\newblock {Estimates of seismic activity in the Cerberus Fossae region of
  Mars}.
\newblock {\em Journal of Geophysical Research: Planets}, 118(12):2570--2581.

\bibitem[Thiriet et~al., 2018]{thiriet18}
Thiriet, M., Michaut, C., Breuer, D., and Plesa, A.-C. (2018).
\newblock Hemispheric dichotomy in lithosphere thickness on mars caused by
  differences in crustal structure and composition.
\newblock {\em J.~Geophys.~Res.}, 123.

\bibitem[Toks{\"o}z et~al., 1974]{toksoz74}
Toks{\"o}z, M.~N., Dainty, A.~M., Solomon, S.~C., and Anderson, K.~R. (1974).
\newblock {Structure of the Moon}.
\newblock {\em Reviews of geophysics}, 12(4):539--567.

\bibitem[Van~Thienen et~al., 2006]{vanThienen06}
Van~Thienen, P., Rivoldini, A., Van~Hoolst, T., and Lognonn{\'e}, P. (2006).
\newblock {A top-down origin for martian mantle plumes}.
\newblock {\em Icarus}, 185(1):197--210.

\bibitem[Vaucher et~al., 2009]{vaucher09}
Vaucher, J., Baratoux, D., Mangold, N., Pinet, P., Kurita, K., and
  Gr{\'e}goire, M. (2009).
\newblock "the volcanic history of central elysium planitia: Implications for
  martian magmatism".
\newblock {\em Icarus}, 204(2):418 -- 442.

\bibitem[Ward, 1998a]{ward98b}
Ward, S.~N. (1998a).
\newblock {On the consistency of earthquake moment rates, geological fault
  data, and space geodetic strain: Europe}.
\newblock {\em Geophys.~J.~Int.}, 135:1011--1018.

\bibitem[Ward, 1998b]{ward98a}
Ward, S.~N. (1998b).
\newblock {On the consistency of earthquake moment rates, geological fault
  data, and space geodetic strain: The United States}.
\newblock {\em Geophys.~J.~Int.}, 134:172--186.

\bibitem[Wieczorek et~al., 2022]{wieczorek22}
Wieczorek, M. et~al. (2022).
\newblock {InSight constraints on the global character of the Martian crust}.
\newblock {\em J.~Geophys.~Res.}

\bibitem[Wieczorek, 2008]{wieczorek08}
Wieczorek, M.~A. (2008).
\newblock {Constraints on the composition of the {Martian} south polar cap from
  gravity and topography}.
\newblock {\em Icarus}, 196(2):506--517.

\bibitem[Wieczorek and Zuber, 2004]{wieczorek04}
Wieczorek, M.~A. and Zuber, M.~T. (2004).
\newblock {Thickness of the martian crust: Improved constraints from
  geoid-to-topography ratios}.
\newblock {\em J.~Geophys.~Res.}, 109 (E1)(E01009):doi:10.1029/2003JE002153.

\bibitem[Wiens and Stein, 1983]{wiens83}
Wiens, D.~A. and Stein, S. (1983).
\newblock {Age dependence of oceanic intraplate seismicity and implications for
  lithospheric evolution}.
\newblock {\em Journal of Geophysical Research: Solid Earth},
  88(B8):6455--6468.

\bibitem[Wilhelms and Squyres, 1984]{wilhelms84}
Wilhelms, D.~E. and Squyres, S.~W. (1984).
\newblock The martian hemispheric dichotomy could be due to a giant impact.
\newblock {\em Nature}, 309:138--140.

\bibitem[Williams and Nimmo, 2004]{williams04}
Williams, J.-P. and Nimmo, F. (2004).
\newblock {Thermal evolution of the Martian core: Implications for an early
  dynamo}.
\newblock {\em Geology}, 32(2):97--100.

\bibitem[Wise et~al., 1979]{wise79}
Wise, D.~U., Golombek, M.~P., and McGill, G.~E. (1979).
\newblock {Tectonic evolution of Mars}.
\newblock {\em J.~Geophys.~Res.}, 84:7934--7939.

\bibitem[Yoshizaki and McDonough, 2020]{yoshizaki20}
Yoshizaki, T. and McDonough, W.~F. (2020).
\newblock {The composition of Mars}.
\newblock {\em Geochimica et Cosmochimica Acta}, 273:137--162.

\bibitem[Zenh{\"a}usern et~al., 2022]{zenhausern22}
Zenh{\"a}usern, G., St{\"a}hler, S.~C., Clinton, J.~F., Giardini, D., Ceylan,
  S., and Garcia, R.~F. (2022).
\newblock {Low-Frequency Marsquakes and Where to Find Them: Back Azimuth
  Determination Using a Polarization Analysis Approach}.
\newblock {\em Bulletin of the Seismological Society of America}.

\bibitem[Zharkov and Gudkova, 2005]{zharkov05}
Zharkov, V. and Gudkova, T. (2005).
\newblock {Construction of Martian interior model}.
\newblock {\em Solar System Research}, 39(5):343--373.

\bibitem[Zheng et~al., 2015]{zheng15}
Zheng, Y., Nimmo, F., and Lay, T. (2015).
\newblock {Seismological implications of a lithospheric low seismic velocity
  zone in Mars}.
\newblock {\em Physics of the Earth and Planetary Interiors}, 240:132--141.

\bibitem[Zhong and Zuber, 2001]{zhong01}
Zhong, S. and Zuber, M.~T. (2001).
\newblock {Degree-1 mantle convection and the crustal dichotomy on Mars}.
\newblock {\em Earth and Planetary Science Letters}, 189:75--84.

\end{thebibliography}

\end{document}